\documentclass[a4paper,english,11pt]{article}

\usepackage{graphicx, amsfonts, amsmath, amssymb, epsfig, color, xparse}
\usepackage{framed}
\usepackage{comment}
\usepackage{caption}

\usepackage[table,xcdraw]{xcolor}

\usepackage[latin9]{inputenc}
\usepackage{babel}
\usepackage{calc}
\usepackage{enumitem}
\usepackage{amsmath}
\usepackage{amsthm}
\usepackage{amssymb}

\usepackage{algorithm,algpseudocode}
\usepackage{amsfonts}
\usepackage{mathtools}
\usepackage[framemethod=TikZ]{mdframed}

\usepackage{fullpage}

\definecolor{armygreen}{rgb}{0.29, 0.33, 0.13}
\definecolor{darkgreen}{rgb}{0.0, 0.5, 0.0}

\newcommand{\BE}{\begin{enumerate}} \newcommand{\EE}{\end{enumerate}}
\newcommand{\BI}{\begin{itemize}} \newcommand{\EI}{\end{itemize}}
\newcommand{\BDes}{\begin{description}}\newcommand{\EDes}{\end{description}}
\newtheorem{alg}{Algorithm}
\newcommand{\BA}{\begin{alg}} \newcommand{\EA}{\end{alg}}

\newcommand{\BEQ}{\begin{equation}} \newcommand{\EEQ}{\end{equation}}
\newcommand{\BEQN}{\begin{eqnarray}}\newcommand{\EEQN}{\end{eqnarray}}

\newcommand{\bitset}{\{0,1\}}

\newcommand{\set}[1]{\left\{ #1 \right\}}

\newcommand{\poly}{{\rm poly}}

\newcommand{\eps}{\epsilon}

\usepackage[colorlinks,citecolor=darkgreen,linkcolor=cyan,bookmarks=true]{hyperref}
\usepackage[nameinlink]{cleveref}

\numberwithin{equation}{section}
\numberwithin{figure}{section}

\newtheorem{theorem}{Theorem}

\newtheorem{lemma}{Lemma}
\newtheorem{claim}[lemma]{Claim}
\newtheorem{observation}[lemma]{Observation}

\newtheorem{proposition}[theorem]{Proposition}

\newtheorem{definition}{Definition}

\theoremstyle{definition}
\newtheorem{step}{Step}

\crefname{substep}{Step}{Steps}

\setlist[enumerate]{itemsep=1pt, parsep=0pt, topsep=2pt, partopsep=0pt}
\setlist[itemize]{itemsep=1pt, parsep=0pt, topsep=2pt, partopsep=0pt}

\makeatletter
\newcommand{\setdef}[1]{%
	\phantomsection
	#1\def\@currentlabel{\unexpanded{#1}}%
}
\makeatother

\makeatletter
\let\ORIGINAL@spythm\@spythm
\def\@spythm#1#2#3#4[#5]{%
	\NR@gettitle{#5}%
	\ORIGINAL@spythm{#1}{#2}{#3}{#4}[#5]%
}
\makeatother

\newcommand{\cycnums}[1]{{\mathbb{Z}_{#1}}}
\NewDocumentCommand{\torus}{mg}{%
	\mathbb{Z}_{#1} \times \mathbb{Z}_{\IfNoValueTF{#2}{#1}{#2}}%
}

\newcommand{\maj}{{\textrm{MAJ}}}

\newcommand{\thr}{{\textrm{Thr}}}

\title{Characterizing and Testing Configuration Stability in Two-Dimensional Threshold Cellular Automata}

\author{
	Yonatan Nakar \\
	\small Tel-Aviv University \\
	\small yonatannakar@mail.tau.ac.il
	\and				
	Dana Ron \\
	\small Tel-Aviv University \\
	\small danaron@tau.ac.il
}

\date{}

\begin{document}
	
\maketitle

\begin{abstract}
	We consider the problems of characterizing and testing the stability of cellular automata configurations that evolve on a two-dimensional torus according to threshold rules with respect to the von-Neumann neighborhood.
	While stable configurations for Threshold-1 (OR) and Threshold-5 (AND) are trivial (and hence easily testable), the other threshold rules exhibit much more diverse behaviors.
	We first characterize the structure of stable configurations with respect to the Threshold-2 (similarly, Threshold-4) and Threshold-3 (Majority) rules. We then design and analyze a testing algorithm that distinguishes between configurations that are stable with respect to the  Threshold-2 rule, and those that are $\epsilon$-far from any stable configuration, where the query complexity of the algorithm is independent of the size of the configuration and depends quadratically on $1/\epsilon$.
\end{abstract}

\thispagestyle{empty}
\newpage
\setcounter{page}{1}

\section{Introduction}
Cellular automata are the building blocks of a wide variety of models and simulations in diverse domains, such as traffic flow~\cite{traffic2022,traffic2011}, population dynamics~\cite{population2011,population2023}, and the propagation of heat~\cite{heat2021,heat1988}, to name a few.
In this paper, we focus on threshold cellular automata that evolve on an $ m \times n $ torus, denoted $ \torus{m}{n} $.
Given a configuration $ \sigma: \torus{m}{n} \to \bitset $ and an integer $ 1 \le b \le 5 $, we denote by $ \thr_b(\sigma) $ the configuration $ \sigma' $ that results from applying the threshold-$ b $ rule on the configuration $ \sigma $.
In this resulting configuration $ \sigma' $, each cell $ \ell \in \torus{m}{n} $ is active -- i.e., satisfies $ \sigma'(\ell) = 1 $ -- if and only if the set $ \Gamma(\ell) \cup \set{\ell} $ includes at least $ b $ cells that were active in $ \sigma $, where $ \Gamma(\ell) $ denotes the von-Neumann Neighborhood of the cell $ \ell $, i.e., $ \Gamma((i,j)) = \set{(i, j \pm 1), (i \pm 1, j)} $.\footnote{
	Observe that the configuration being tordial ensures that $ \Gamma(\ell) $ is well-defined for every cell $ \ell $.
	An alternative setting is non-cyclical configurations, with the assumption of a $ 0 $ or $ 1 $ fixed-state boundary.
	We note that the latter can be seamlessly reduced to the former, and hence the former is preferable.
}

Two-dimensional threshold rules are simple.
Yet, despite their apparent simplicity, certain rules -- specifically, the Majority rule (Threshold-$ 3 $), the Threshold-$ 2 $ rule and its analog, Threshold-$ 4 $ -- can produce a broad array of complex behaviors, prompting various questions about their dynamics.
Interestingly, even with the richness of their dynamics, every threshold cellular automaton, regardless of its initial configuration, is guaranteed to ultimately converge to either a fixed-point configuration or to a cycle comprising a pair of configurations~\cite{goles1981_period_2}.
We refer to such ``final'' configurations as \emph{stable}.
That is, a configuration $ \sigma: \torus{m}{n} \to \bitset $ is stable with respect to the Threshold-$ b $ rule if $ \thr_b^2(\sigma) = \sigma $, where $ \thr_b^2(\sigma) = \thr_b(\thr_b(\sigma)) $.

In this work, we study the question of configuration stability under the paradigm of property testing~(see, e.g., the textbooks \cite{odedbook,yuichibook}).
Namely, given query access to a configuration $ \sigma: \torus{m}{n} \to \bitset $, we would like to determine (with high constant success probability) whether the configuration $ \sigma $ is stable or whether it significantly deviates from being stable.
The measure of this deviation is derived from the normalized Hamming distance between pairs of configurations, where $ \sigma $ is considered $ \eps $-far from being stable if more than an $ \eps $-fraction of the cells must be modified in order to obtain a stable configuration.
The testing algorithm's query complexity and running time should ideally be polynomial in $ 1/\eps $ and independent of $ m $ and $ n $.

The problem of testing configuration stability under any two-dimensional threshold rule can be reduced to the problem of testing local forbidden patterns in two-dimensional matrices, as studied in \cite{k_local}.\footnote{
	Even though the testing algorithm from \cite{k_local} is designed for matrices, it is adaptable to torii.
}
In particular, since every stable configuration $ \sigma : \torus{m}{n} \to \bitset $ satisfies $ \thr_b^2(\sigma) = \sigma $, and since the next state of each cell is a function of the cell's von-Neumann neighborhood, the problem of testing configuration stability is equivalent to being free of a family of $ 5 \times 5 $ unstable sub-matrices that depends on the specific threshold rule.
In \cite{k_local}, the author presents a testing algorithm for the more general problem of pattern freeness in $ d $-dimensional arrays over a finite alphabet.
The query complexity of the testing algorithm is linear in $ m $ and $ n $ (and the running time is exponential).

We ask whether it is possible to test configuration stability under two-dimensional threshold rules with query complexity (and running time) that have a much better dependence on $ m $ and $ n $, and possibly with no dependence at all.
A naive algorithm for testing configuration stability is selecting, uniformly and independently at random, a sample $ \mathcal{L} \subseteq \torus{m}{n} $ of cells, and accepting if and only if every cell $ \ell \in \mathcal{L} $ satisfies $ \sigma''(\ell) = \sigma(\ell) $, where $ \sigma'' = \thr_b^2(\sigma) $.
However, we show in \Cref{sec:naive}, that, for the Threshold-$ 2 $, $ 3 $ and $ 4 $ rules, the naive algorithm requires a sample size that depends on the configuration size, and the examples we provide there suggest that, to improve upon the naive approach, we need an algorithm that takes into account the ``non-local'' patterns in the structure of stable configurations.

Thus, in order to design more efficient algorithms for testing configuration stability, our approach is to first uncover the structure of stable configurations for the given threshold rule.
For the OR (Threshold-$ 1 $) and AND (Threshold-$ 5 $) rules, the stable configurations are the all-$ 0 $s and the all-$ 1 $s configurations, so the task of testing configuration stability is trivial under these two rules.
In this paper, we provide a characterization of the structure of the Threshold-$ 2 $ (similarly, Threshold-$ 4 $) stable configurations, as well as that of the Majority rule.
Based on the characterization of the former, we design a one-sided error testing algorithm for configuration stability under the Threshold-$ 2 $ rule (and hence also Threshold-$ 4 $), whose query complexity is independent of the configuration's dimensions and is polynomial in $ 1/ \eps $.
This algorithm searches for several types of violations of constraints imposed by the characterization (with different ``levels of locality'').

As can be seen from the statements of the characterizations (\Cref{claim:limit_cycles_in_th2} and \Cref{claim:majority_strongly_stable_configurations}), stable configurations of the Majority rule have structures that are quite a bit more intricate than those of Threshold-$ 2 $ rule.
This suggests that the task of designing an efficient testing algorithm for configuration stability under the Majority rule is more complex.
This problem indeed remains open and we hope that the characterization we provide will serve as a building block for resolving it.

\subsection{Our results}
Our main contribution is a testing algorithm for configuration stability under the Threshold-$ 2 $ rule.
Since the Threshold-$ 4 $ rule operates identically to the Threshold-$ 2 $ rule when the roles of $ 0 $s and $ 1 $s are reversed, the same result holds for the Threshold-$ 4 $ rule as well.

\begin{theorem}\label{thm:testable}
	There exists a one-sided error testing algorithm with query complexity $ O(1/\eps^2) $ for testing configuration stability under the two-dimensional Threshold-$ 2 $ rule.
\end{theorem}

The testing algorithm and its analysis appear in \Cref{sec:testing_alg}.
As mentioned, we have two structural results, one for the Threshold-$ 2 $ rule and another for the Majority rule.
Both structural results are based on the existence of a \emph{partition} of the torus into what we refer to as monochromatic and chessboard sets.\footnote{Another type of structure we call \nameref*{def:zebra}s is needed only for the majority rule.}
Formally,
\begin{definition}[monochromatic set]
	Given a configuration $ \sigma : \torus{m}{n} \to \bitset $ and a value $ \beta \in \bitset $, we say that a connected set of cells $ C \subseteq \torus{m}{n} $ is \textsf{$ \beta $-monochromatic} with respect to $ \sigma $ if $ \sigma(\ell) = \beta $ for every cell $ \ell \in C $.
\end{definition}

\begin{definition}[chessboard set]
	Given a configuration $ \sigma : \torus{m}{n} \to \bitset $, we say that a connected set of cells $ C \subseteq \torus{m}{n} $ is a \textsf{chessboard set} if $ |C| \ge 2 $ and if $ \sigma(\ell_1) \ne \sigma(\ell_2) $ for every pair of adjacent cells $ \ell_1, \ell_2 \in C $.
\end{definition}

The difference between our two structural results, the one for the Threshold-$ 2 $ rule versus the one for the Majority rule, lies in the particular constraints that the partition must satisfy.

\subsubsection{Threshold-2 rule structural result}
In the case of the Threshold-$ 2 $ rule, one of the constraints is that the sets are not just any monochromatic or chessboard sets, but what we refer to as rectangular \emph{\nameref*{def:component}s}.

\begin{definition}[monochromatic component]
	We say that a set $ C \subseteq \torus{m}{n} $ is a \textsf{\nameref*{def:mono_component}} if $ C $ is a maximal monochromatic set and $ |C| \ge 2 $.
\end{definition}

\begin{definition}[chessboard component]
	We say that a set $ C \subseteq \torus{m}{n} $ is a \textsf{\nameref*{def:chess_component}} if $ C $ is a chessboard set such that every cell $ \ell \in C $ is adjacent to at least two other cells in $ C $, and, furthermore $ C $ is maximal with respect to this requirement.
\end{definition}

Moreover, we say that a chessboard or monochromatic \nameref*{def:component} $ C \in \torus{m}{n} $ is a \emph{rectangular \nameref*{def:component}} if there is a pair of intervals $ I \subseteq \cycnums{m} $ and $ J \subseteq \cycnums{n} $ (possibly wrapping around) such that $ C = \set{(i,j) \in \torus{m}{n} | i \in I, j \in J} $.

Since the set of stable configurations is characterized by partitions of the \emph{torus}, those partitions can potentially contain chessboard or monochromatic \nameref*{def:component}s that are \emph{wraparound rectangles}.
However, as we'll see, no \nameref*{def:mono_component} in these partitions can be an \emph{almost-wraparound rectangle}.

\begin{definition}[(almost-)wraparound rectangle]
	A rectangle $ R = I \times J $, where $ I \subseteq \cycnums{m} $ and $ J \subseteq \cycnums{n} $ are a pair of coordinate intervals, is said to be a \textsf{wraparound rectangle} if either $ I = \cycnums{m} $ or $ J = \cycnums{n} $ (or both).
	If either $ I = \cycnums{m} \setminus \set{i} $ for some $ i \in \cycnums{m} $ or $ J = \cycnums{n} \setminus \set{j} $ for some $ j \in \cycnums{n} $ (or both), then $ R $ is an \textsf{almost-wraparound rectangle}.
\end{definition}

\begin{figure}
	\centering
	\begin{minipage}{0.46\textwidth}
		\centering
		\includegraphics[width=.95\textwidth]{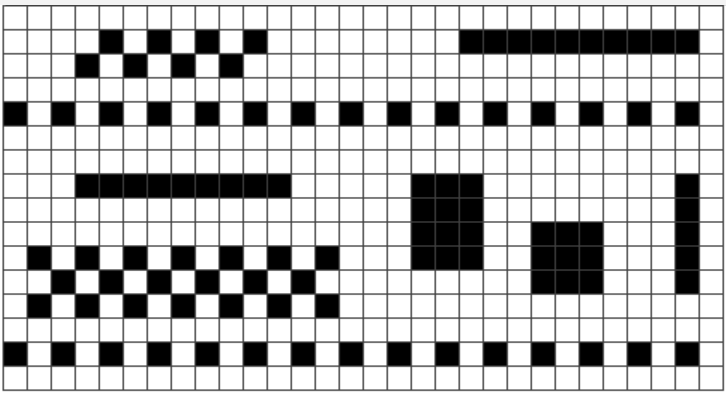}
	\end{minipage}
	\begin{minipage}{0.46\textwidth}
		\centering
		\includegraphics[width=.95\textwidth]{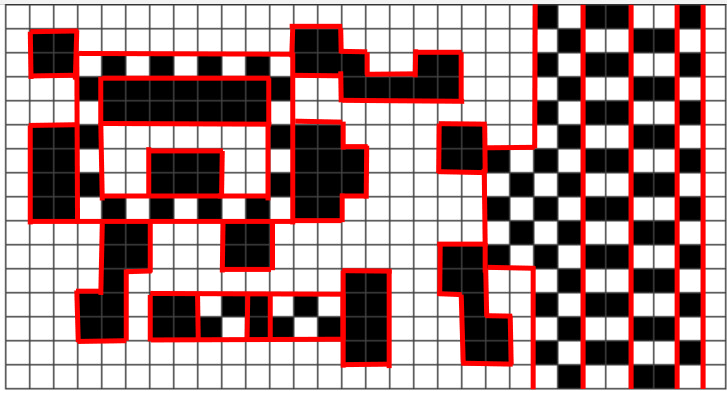}
	\end{minipage}
	\caption{\small
		\textbf{Left:} An example of a stable configuration for the Threshold-2 rule (where the illustrated matrix represents a torus, i.e., the first and last rows/columns are adjacent).
		Dark cells correspond to state-$ 1 $ cells and white cells correspond to state-$ 0 $ cells.
		The configuration contains four chessboard \nameref*{def:component}s, two of which are \nameref*{def:wraparound} components, and five monochromatic components.
		Each chessboard \nameref*{def:component} is toggling and each monochromatic \nameref*{def:component} is fixed.
		Since all the \nameref*{def:component}s are rectangular and the distance requirements hold, the configuration is indeed stable.
		\textbf{Right:} An example of a stable configuration for the majority rule.
		The boundaries between the different sets in the partition are outlined in red.
		There are five chessboard sets, thirteen dark monochromatic sets, two white monochromatic sets and two \nameref*{def:zebra}s.
		The chessboard sets and the \nameref*{def:zebra}s are toggling, and the monochromatic sets are fixed.
	}\label{fig:stable_configurations}
\end{figure}

We next state the characterization itself (\Cref{claim:limit_cycles_in_th2}).
For an illustration, see \Cref{fig:stable_configurations}.
As we explain in \Cref{sec:related_work}, some aspects of this characterization appear in~\cite[Proposition 4.5]{goles2013neural}, but a precise characterization and its proof were not provided there.

\begin{proposition}\label{claim:limit_cycles_in_th2}
	Let $ \sigma : \torus{m}{n} \to \bitset $ be a configuration.
	Denote by $ \mathcal{X} \subseteq 2^{\torus{m}{n}} $ the set of all state-1 \nameref*{def:mono_component}s in $ \sigma $ and by $ \mathcal{Y} \subseteq 2^{\torus{m}{n}} $ the set of all \nameref*{def:chess_component}s in $ \sigma $.
	Let $ X $, $ Y $ and $ Z $ be the following sets of cells.
	\begin{align*}
		X = \bigcup_{C \in \mathcal{X}} C, \qquad
		Y = \bigcup_{C \in \mathcal{Y}} C, \qquad
		Z = (\torus{m}{n}) \setminus (X \cup Y) .
	\end{align*}
	The configuration $ \sigma $ is stable with respect to the Threshold-$ 2 $ rule if and only if all of the following hold.
	\begin{enumerate}
		\item\label{item:all_components_are_rectangles} Every set $ C \in \mathcal{X} \cup \mathcal{Y} $ is a rectangle and no set $ C \in \mathcal{X} $ is an almost-wraparound rectangle.
		\item\label{item:everything_outside_components_zero} Every cell $ \ell \in Z $ satisfies $ \sigma(\ell) = 0 $.
		\item\label{item:rectangles_not_too_close} For every pair of distinct sets $ R,R' \in \mathcal{X} \cup \mathcal{Y} $, the Manhattan distance between $ R $ and $ R' $ is at least 2, and if either $ R $ or $ R' $ belong to $ \mathcal{X} $, then the Manhattan distance is at least 3.
	\end{enumerate}
\end{proposition}

We note that in stable configurations, all the cells $ \ell $ that belong to \nameref*{def:chess_component}s are \emph{toggling} (i.e., satisfy $ \sigma'(\ell) \ne \sigma(\ell) $ where $ \sigma' = \thr_b(\sigma) $) and all other cells are \emph{fixed} (i.e., satisfy $ \sigma'(\ell) = \sigma(\ell) $).
The proof of \Cref{claim:limit_cycles_in_th2} appears in \Cref{sec:threshold2_structure_proof}.

\subsubsection{Majority rule (Threshold-3) structural result}
The characterization of stable configurations for the Majority rule is somewhat more complex than the one for the Threshold-2 rule.
On a high level, in stable configurations, the cells can be partitioned into four collections of connected sets.
The first collection contains $0$-monochromatic sets, the second contains $1$-monochromatic sets, the third contains chessboard sets and the fourth contains sets of a type that we haven't yet introduced (which we name ``\nameref*{def:zebra}'' -- see \Cref{def:zebra}).
In contrast to the Threshold-2 rule, the monochromatic and chessboard sets are not necessarily rectangular, but may have varying structures. Furthermore, there are no constraints on the distances between them.
However, there are constraints on neighboring sets.
For example, if a cell belongs to a $1$-monochromatic set that has three neighboring cells outside of it, two of these neighbors must belong to chessboard sets, and they must be of opposite states.
For an illustration, see \Cref{fig:stable_configurations}.

\begin{definition}[zebra wraparound rectangle]\label{def:zebra}
	Let $ \sigma: \torus{m}{n} \to \bitset $ be a configuration and let $ k $ be an integer. We say that a set $ C \subset \torus{m}{n} $ is a \textsf{width-$ k $ \nameref*{def:zebra}} if it satisfies one of the following conditions:
	
	\begin{enumerate}
		\item $ C $ is an $ m \times k $ wraparound rectangle, where every row of $ C $ forms a monochromatic set in $ \sigma $ and every column of $ C $ forms a chessboard set in $ \sigma $.
		
		\item $ C $ is a $ k \times n $ wraparound rectangle, where every column of $ C $ forms a monochromatic set in $ \sigma $ and every row of $ C $ forms a chessboard set in $ \sigma $.
	\end{enumerate}
\end{definition}

We next state the Majority rule structural result (\Cref{claim:majority_strongly_stable_configurations}).
The proof can be found in \Cref{sec:majority_structure_proof}.

\begin{proposition}\label{claim:majority_strongly_stable_configurations}
	A configuration $ \sigma : \torus{m}{n} \to \bitset $ is stable with respect to the Majority rule if and only if there is a partition $ \mathcal{P} $ of $ \torus{m}{n} $ into connected sets of cells that satisfies the following requirements:
	\begin{enumerate}
		\item\label{req:components} $ \mathcal{P} = \mathcal{P}_0 \cup \mathcal{P}_1 \cup \mathcal{P}_2 \cup \mathcal{P}_3 $ where all the sets in $ \mathcal{P}_0 $ are $ 0 $-monochromatic sets, all the sets in $ \mathcal{P}_1 $ are $ 1 $-monochromatic sets, all the sets in $ \mathcal{P}_2 $ are chessboard sets, and all the sets in $ \mathcal{P}_3 $ are width-$ 2 $ \nameref*{def:zebra}s.
		
		\item\label{req:adjacency} For each $ k \in \set{0,1,2} $, no two sets in $ \mathcal{P}_{k} $ are adjacent to each other.
		
		\item\label{req:mono} If $ C \in \mathcal{P}_0 \cup \mathcal{P}_1 $, then for every cell $ \ell \in C $, if $ |\Gamma(\ell) \setminus C| = 3 $, there are two cells $ \ell_0, \ell_1 \in \Gamma(\ell) \cap \bigcup_{C' \in \mathcal{P}_2} C' $ satisfying $ \sigma(\ell_0) = 0 $ and $ \sigma(\ell_1) = 1 $, and if $ |\Gamma(\ell) \setminus C| = 4 $, there are exactly two cells $ \ell_0', \ell_0'' \in \Gamma(\ell) \cap \bigcup_{C' \in \mathcal{P}_2} C' $ and exactly two cells $ \ell_1', \ell_1'' \in \Gamma(\ell) \cap \bigcup_{C' \in \mathcal{P}_2} C' $ satisfying $ \sigma(\ell_0') = \sigma(\ell_0'') = 0 $ and $ \sigma(\ell_1') = \sigma(\ell_1'') = 1 $.
		
		\item\label{req:chess} If $ C \in \mathcal{P}_2 $, then every cell $ \ell \in C $ is adjacent to at most one cell belonging to $ \bigcup_{C' \in \mathcal{P}_0} C' $ and to at most one cell belonging to $ \bigcup_{C' \in \mathcal{P}_1} C' $.
		
		\item\label{req:zebra} If $ C \in \mathcal{P}_3 $, then every cell in $ \ell \in C $ is adjacent to a cell $ \ell' \in C' $ for some $ C' \in \mathcal{P}_2 \cap \mathcal{P}_3 \setminus \set{C} $ where $ \sigma(\ell) \ne \sigma(\ell') $.
	\end{enumerate}
\end{proposition}

\subsection{An overview of the Threshold-2 testing algorithm and its analysis}\label{sec:high_level}
As stated in \Cref{claim:limit_cycles_in_th2}, a configuration $\sigma$ is stable if and only if the monochromatic and chessboard \nameref*{def:component}s of $\sigma$ are rectangular submatrices (where the monochromatic ones are not almost-wraparound rectangles), and there are certain constraints on the minimum distance between the rectangles.
One  implication of the definition of \nameref*{def:chess_component}s is that a stable configuration may contain a chessboard rectangle of height/width $ 1 $ only if it is a wraparound rectangle.
On the other hand, a monochromatic rectangle in a stable configuration may have height or width $ 1 $, though not both.

\smallskip\noindent
\textbf{A simplifying assumption.}~
Assume first, for the sake of the exposition, that stable configurations allowed for chessboard rectangles with height/width $ 1 $ as well as for monochromatic rectangles of size $ 1 $.
We later explain how to remove this  assumption.
For a parameter $k = c_1/\epsilon$, where $c_1$ is some (sufficiently large) constant, consider performing the following ``gridding''/``rectangulation'' operation on $\sigma$: for every $k$th row, we replace the row and each of the two adjacent rows by an all-$0$ row, and the same process is applied to every $k$th column.

Let the resulting configuration be denoted $\sigma^\#$.
Observe that by our (incorrect, but simplifying) assumption, if $\sigma$ is stable, then so is $\sigma^\#$.
Moreover, if $\sigma$ is $\eps$-far from being stable, then $\sigma^\#$ is $(\eps/2)$-far from stable (for $c_1 \geq 8$).
The benefit of considering $\sigma^\#$ is that it allows the testing algorithm to be more local and attempt to find evidence that the configuration is not stable within the $k\times k$ rectangles defined by $\sigma^\#$.

Specifically, the algorithm takes a sample of $\Theta(1/\eps)$ cells.
For each selected cell it queries its local (distance-$3$) neighborhood.
If it finds evidence against stability in the local neighborhood, then it rejects.
Otherwise, the cell is locally consistent with belonging to either a monochromatic rectangle or a chessboard rectangle, and the algorithm essentially tries to determine this rectangle in $\sigma^\#$ (assuming $\sigma$, and hence $\sigma^\#$, is stable).
 
We next describe how this is done for a state-$1$ selected cell $\ell$ in monochromatic components (the procedure for a cell belonging to a chessboard components is similar).
The algorithm queries $\sigma$ to find the longest paths in $\sigma^\#$ of consecutive state-$1$ cells starting from $\ell$ and going to the right, to the left, up and down.
This creates a cross-shaped path of state-$1$ cells, which in turn defines a rectangle that we refer to as the \emph{bounding box} of $\ell$ (for an illustration, see \Cref{fig:bounding_box}).

Next, the algorithm takes a sample of $\Theta(1/\epsilon)$ cells  within the defined bounding box and checks that all sampled cells are state-$1$ cells.
In addition, it checks that all $O(k) = O(1/\eps)$ cells on the internal \nameref*{def:boundary} of the bounding box have state $1$, and that all cells at distance $ 1 $ and $ 2 $ from the bounding box have state $0$ (in $\sigma^\#$).
If any of these checks fails, then the algorithm rejects. Otherwise, it accepts.

\smallskip\noindent
\textbf{The analysis under the simplifying assumption.}~
It is not hard to verify that if $\sigma$ is stable, then the algorithm always accepts (since for any selected cell $\ell$, the algorithm determines precisely its rectangular component in $\sigma^\#$).
The crux of the proof is showing that if $\sigma$ is $\eps$-far from being stable (so that $\sigma^\#$ is $\eps/2$-far from being stable), then the algorithm rejects with probability at least $2/3$.
To this end we prove the contrapositive statement.
Namely, that if the algorithm rejects with probability smaller than $2/3$, then this implies that $\sigma^\#$ is $\eps/2$-close to being stable.
This is done by ``stabilizing'' $\sigma^\#$, as we explain next.

We say that a bounding box of a cell $\ell$ is \emph{$\alpha$-good}, if the boundary conditions defined above hold perfectly, and the fraction of cells within the bounding box that would have caused the algorithm to reject had it selected $\ell$, is at most $\alpha$.
We refer to the latter as \emph{violating} cells with respect to the bounding box of $\ell$ (for an illustration, see \Cref{fig:violations}).
A simple but useful observation is that for any $\alpha < 1$, due to the boundary conditions, any two $\alpha$-good bounding boxs are either disjoint, or one is contained in the other.
This is easy to verify for the monochromatic case, but also holds for chessboard bounding boxes.

Given this observations, setting $\alpha = \epsilon/c_2$ (for a sufficiently large constant $c_2$), the stabilization procedure considers all maximal $\alpha$-good bounding boxes, and fixes the violations within them.
Since the bounding boxes are disjoint, the total number of modifications performed is at most $ \alpha mn \le \eps mn / 4 $.
In addition, all $1$-cells that do not belong to any maximal $\alpha$-good bounding box are set to $0$. Conditioned on the algorithm rejecting with probability smaller than $2/3$, the total fraction of such cells must be at most $\eps/4$.

\smallskip\noindent
\textbf{Removing the assumption.}~
It remains to discuss how to modify the algorithm and its analysis based on the actual characterization of stable configurations, which may contain rectangular \nameref*{def:chess_component}s of height/width $ 1 $ only if they wrap around, and any \nameref*{def:mono_component} must be of size at least $ 2 $.
In what follows, for the sake of conciseness, we refer only to \nameref*{def:wraparound} rows, rather than rows or columns.

The main modification in the algorithm is an additional initial step which tries to catch rows that contain relatively many cells that ``potentially'' belong to a chessboard \nameref*{def:wraparound} row but whose row is actually relatively far from being such a wraparound row.
By ``potentially'' belonging we mean that their local neighborhood is consistent with belonging to such a row.
Roughly speaking, this step causes rejection if it detects in a certain row one cell that potentially belongs to a chessboard wraparound row and another that is inconsistent with such a row.
The second step of the algorithm is essentially the same as under the simplifying assumption, except that $\sigma^\#$ is defined slightly differently, and sampled cells that potentially belong to chessboard \nameref*{def:wraparound} rows are ignored.

Turning to the analysis, here too we prove that either the algorithm rejects with probability at least $2/3$, or $\sigma$ is $\eps$-close to being stable.
To this end we also define a stabilization procedure.
This procedure first fixes rows (or columns) that are close to being chessboard \nameref*{def:wraparound}.
This means modifying cells in these rows as well as rows at distance $ 1 $ and $ 2 $ from them.
We show that if the algorithm does not reject in its first step with probability at least $2/3$, then the total number of these modifications is $O(\eps mn)$.

Next the stabilizing procedure sets to $0$ those cells that are $1$ in $\sigma$ and $0$ in $\sigma^\#$ \emph{except} if they belong to a row that was fixed to become wraparound (so as not to ``ruin'' wraparound rows).
In its third step the procedure fixes maximal $\alpha$-good bounding boxes.
If they do not intersect any wraparound row, then this is done as defined under the simplifying assumption.
Otherwise, the fixing procedure is a bit more complex.
Finally, it turns to $0$ all state-$1$ cells that do not belong to any \nameref*{def:wraparound} row nor to any maximal $\alpha$-good bounding box.

\subsection{Related Work}\label{sec:related_work}
Testing configuration stability can serve as a building block for testing dynamic environments as studied in \cite{GR-dyn,testing_1d_rules,testing_OR_arbitrary_graphs}.
In the problem of testing dynamic environments, as defined in \cite{GR-dyn}, we are given query access to a \emph{sequence} of configurations, and have to distinguish between the case in which the sequence corresponds to an evolution that obeys a prespecified rule and the case in which the sequence is far from obeying the rule.
Dynamic environments that evolve according to threshold rules always ultimately converge to a cycle of at most two configurations within a number of steps that is linear in the configuration size~\cite{goles1981_period_2}.
Therefore, if the length of the sequence is much larger than the configuration size, the problem of testing dynamic environments reduces to the problem of testing configuration stability.

Testing dynamic environments that evolve according to one-dimensional threshold rules was studied in \cite{testing_1d_rules}.
The structure of stable configurations in one-dimensional threshold rules, in contrast to that of two-dimensional threshold rules, is relatively simple, which makes the problem of testing configuration stability in one-dimensional threshold rules rather trivial.
However, if we increase the size of the local neighborhoods defining the rule, then even in the one-dimensional case, the structure of stable configurations becomes somewhat more complex, as shown in the characterization provided in~\cite{1d_maj_structure}.
Nevertheless, \cite{1d_maj_structure}'s characterization can be used to design a simple $ \poly(1/\eps) $ ``enforce and test'' algorithm for configuration stability of the one-dimensional majority rule for a general neighborhood size.
A further possible generalization of the neighborhood relationship can be defined by an arbitrary underlying graph.
This generalization was recently studied by Modanese and Yoshida \cite{testing_OR_arbitrary_graphs} in the context of testing dynamic environments where the evolution rule is Threshold-$ 1 $ (the OR rule).
We note that in this setting, the problem of testing configuration stability, unlike the problem of testing dynamic environments, is easy to solve, because stable configurations are characterized by all vertices of each connected component in the underlying graph having the same state.

A simplified version of our testing algorithm can be used to test the following  property of binary matrices: The $1$-entries of the matrix can be partitioned into disjoint (contiguous) rectangles of size at least $2$ such that every two rectangles are at distance at least $2$ from each other.\footnote{Both the minimum size of the rectangles and the distances can be replaced by other constants.}
Two related property testing problems that have been studied previously~\cite{Bsh,PRS} are whether the matrix has binary rank at most $d$ and whether it has Boolean rank at most $d$.
The former corresponds to there being a partition of the $1$-entries of the matrix into at most $d$ disjoint \emph{generalized} rectangles, and the latter to there being a cover of the $1$-entries of the matrix by at most $d$ generalized rectangles.

As mentioned earlier, there exists a testing algorithm for the more general problem of pattern freeness in multi-dimensional arrays over a finite alphabet, which can be used to test configuration stability of two-dimensional threshold rules, but the algorithm's query complexity is linear in $ m $ and $ n $~\cite{k_local}.
There is another result for the problem of pattern freeness in multi-dimensional arrays over a finite alphabet where the query complexity is linear in $ 1/\eps $ and does not depend on the dimensions of the array~\cite{testing_pattern_freeness}.
However, that testing algorithm is applicable to the property of freeness from a \emph{single} pattern, while the property of configuration stability corresponds to freeness from a \emph{family} of patterns.
Moreover, for both the Threshold-$ 2 $ and the majority rule, some of the patterns in the family of patterns are what the authors of \cite{testing_pattern_freeness} refer to as almost-homogeneous patterns, to which their approach does not apply either.

Other results regarding the structure of stable configurations for threshold rules include a characterization of the structure of stable configurations in majority processes evolving on trees~\cite{trees_turau2022}, as well as results related to conditions under which a random initial configuration ultimately converges to a monochromatic configuration.
Among these are majority processes evolving on random regular graphs~\cite{Zehmakan2018majority}, on Erdos-Renyi random graphs~\cite{zehmakan2020opinion}, on expanders~\cite{zehmakan2020opinion}, and, finally, on the two-dimensional torus~\cite{Zehmakan2021majority_2d_cellular_automata}.

In \cite[Chap. 4]{goles2013neural}, there is an investigation of the structure of stable configurations for two-dimensional threshold rules on an infinite grid with a finite number of state-$ 1 $ cells.
The investigation is undertaken for the purpose of bounding the number of steps until a stable configuration is reached.
They characterize the structure of the fixed-point configurations of the Threshold-$ 2 $ rule in this setting as well as the structure of stable configurations in which all the state-$ 1 $ cells are toggling.
Regarding the two-dimensional majority rule, they do not provide a characterization of the set of stable configurations, but they do prove some properties of stable configurations (such as the existence of certain forbidden patterns).
\section{Testing configuration stability under the Threshold-2 rule}\label{sec:testing_alg}
In this section we present a one-sided error testing algorithm that, given query access to a configuration $ \sigma: \torus{m}{n} \to \bitset $, \textsf{accepts} $ \sigma $ if it is stable under the Threshold-2 rule, and \textsf{rejects} $ \sigma $ with high constant probability if it is $ \eps $-far from being stable.

\begin{definition}\label{def:stable}
	Given a configuration $ \sigma: \torus{m}{n} \to \bitset $ and a cell $ \ell \in \torus{m}{n} $, we say that the cell $ \ell $ is \textsf{stable} with respect to the Threshold-$ b $ rule if $ \sigma''(\ell) = \sigma(\ell) $, where $ \sigma'' = \thr_b^2(\sigma) $.
\end{definition}

Given a set of cells $ X \subseteq \torus{m}{n} $, we denote by $ \sigma[X] $ the sub-configuration of $ \sigma $ induced by $ X $.
Additionally, given a cell $ (i,j) \in \torus{m}{n} $ and an integer $ r \ge 0 $, we define its \emph{distance-$ r $ neighborhood}, denoted $ \Gamma_{\le r}((i,j)) $, as the set of cells at Manhattan distance \emph{at most} $ r $ from $ (i,j) $.
We also denote by $ \Gamma_{=r}((i, j)) $ the set of cells at distance \emph{exactly} $ r $ from $ (i,j) $.

In a nutshell, the algorithm looks for three kinds of violations.
The first kind has a global nature, involving chessboard wraparound components.
The second is of a moderate locality, involving sampling from $ O(1/\eps) $-neighborhoods of cells.
The third is a simple violation of local stability.
We begin with the first.

\subsection{Wraparound violating pairs}
The notion of \nameref*{def:wraparound_violating_pair}s arises from noticing that, in stable configurations, \nameref*{def:chess_component}s cannot be subsets of \emph{single} rows or columns, unless they occupy the entire row or column.
That is, these \nameref*{def:component}s must be \emph{chessboard \nameref*{def:wraparound}s}.

\begin{definition}[wraparound]\label{def:wraparound}
	Given a configuration $ \sigma: \torus{m}{n} $, we say that row $ r \in \cycnums{m} $ (column $ c \in \cycnums{n} $) is an \textsf{even chessboard \nameref*{def:wraparound}} if all of the following hold:
	\begin{enumerate}
		\item The entire row (column) is a \nameref*{def:chess_component} in $ \sigma $.
		
		\item Rows $ r-1 $ and $ r+1 $ (columns $ c-1 $ and $ c+1 $) consist entirely of state-$ 0 $ cells.
		
		\item No \nameref*{def:mono_component} is at distance less than $ 3 $ from row $ r $ (column $ c $).
		
		\item\label{item:even_position} Every state-$ 1 $ cell in row $ r $ (column $ c $) is positioned at an even-numbered column (row).
	\end{enumerate}
	Similarly, we say that row $ r \in \cycnums{m} $ (column $ c \in \cycnums{n} $) is an \emph{odd} chessboard \nameref*{def:wraparound} if all of the above hold but requirement~\ref{item:even_position} is replaced by the requirement that every state-$ 1 $ cell in row $ r $ (column $ c $) is positioned at an \emph{odd}-numbered column (row). 
\end{definition}

We extend the notion of \nameref*{def:wraparound} rows and columns by introducing an approximate variant termed \emph{$ \alpha $-\nameref*{def:wraparound_consistent}}.

\begin{definition}[wraparound consistent]\label{def:wraparound_consistent}
	Given a configuration $ \sigma: \torus{m}{n} \to \bitset $, we say that a cell $ (r,c) \in \torus{m}{n} $ is \textsf{row (column) even (odd) chessboard \nameref*{def:wraparound_consistent}} if $ \sigma[\Gamma_{\le 3}(\ell)] $ is \textsf{consistent with} row $ r $ (column $ c $) being an \textsf{even (odd) chessboard \nameref*{def:wraparound}} (implicitly implying the row's (column's) length must be even).
	That is, if there exists a configuration $ \sigma': \torus{m}{n} \to \bitset $ that agrees with $ \sigma $ on every cell $ \ell' \in \Gamma_{\le 3}(\ell) $ such that row $ r $ (column $ c $) is an even (odd) chessboard \nameref*{def:wraparound} in $ \sigma $.
	
	We say that a row $ r \in \cycnums{m} $ (column $ c \in \cycnums{n} $) is even (odd) $ \alpha $-chessboard \nameref*{def:wraparound_consistent} for some $ 0 < \alpha < 1 $, if at least $ 1-\alpha $ fraction of its cells are row (column) even (odd) chessboard \nameref*{def:wraparound_consistent}.
\end{definition}

Every chessboard \nameref*{def:wraparound_consistent} cell in a stable configuration $ \sigma $ ``enforces'' an entire row or an entire column to be either even or odd \nameref*{def:wraparound_consistent}.
Moreover, the existence of a cell that is row (column) chessboard \nameref*{def:wraparound_consistent}, rules out the existence of chessboard \nameref*{def:wraparound_consistent} columns (rows) in the entire configuration.
These observations are the basis of the notion of \nameref*{def:wraparound_violating_pair}s.

\begin{definition}[wraparound violating pair]\label{def:wraparound_violating_pair}
	Given a configuration $ \sigma : \torus{m}{n} $, we say that a pair of cells $ \ell, \ell' \in \torus{m}{n} $ constitutes a \textsf{\nameref*{def:wraparound_violating_pair}} if one of the following holds:
	\begin{enumerate}
		\item Both $ \ell $ and $ \ell' $ belong to the same \textsf{row}, one of them is row-even (odd) chessboard \nameref*{def:wraparound_consistent} and the other is either row-odd (even) chessboard \nameref*{def:wraparound_consistent} or not row chessboard \nameref*{def:wraparound_consistent} at all.
		
		\item Both $ \ell $ and $ \ell' $ belong to the same \textsf{column}, one of them is column-even (odd) chessboard \nameref*{def:wraparound_consistent} and the other is either column-odd (even) chessboard \nameref*{def:wraparound_consistent} or not column chessboard \nameref*{def:wraparound_consistent} at all.
		
		\item One of the two cells is \textsf{row} chessboard \nameref*{def:wraparound_consistent} and the other is \textsf{column} chessboard \nameref*{def:wraparound_consistent}.
	\end{enumerate}
\end{definition}

\subsection{The bounding box of the \nameref*{def:plus}}
Our testing algorithm queries the \nameref*{def:plus} of each cell in the algorithm's sample, as well as the perimeter of the \nameref*{def:plus}'s bounding box.
In order to formally define those terms, we have to use the cell's \nameref*{def:moore}, rather than its von-Neumann Neighborhood, as well as what we refer to as \nameref*{def:moore} $ 1 $-\nameref*{def:isolated} cells.

\begin{definition}[Moore neighborhood]\label{def:moore}
	Given a cell $ (i,j) \in \torus{m}{n} $, the cell's \textsf{\nameref*{def:moore}} is composed of the cell $ (i,j) $ and the eight cells that surround it.
	That is, the \nameref*{def:moore} of the cell $ (i,j) $ is the set $ \set{(i+k,j+k) \; \colon -1 \le k \le 1} $.
\end{definition}

\begin{definition}[isolated]\label{def:isolated}
	Given a configuration $ \sigma: \torus{m}{n} \to \bitset $ and a cell $ \ell \in \torus{m}{n} $, we say that the cell $ \ell $ is a \textsf{\nameref*{def:moore} $ 1 $-\nameref*{def:isolated} cell} if $ \sigma(\ell) = 1 $ and every cell $ \ell' \ne \ell $ in the \nameref*{def:moore} of $ \ell $ satisfies $ \sigma(\ell') = 0 $.
\end{definition}

The definition of a cell's \nameref*{def:plus} (defined below) depends on whether the cell is a monochromatic or a chessboard cell.
We have defined these terms only for sets of cells, but there is a natural variant of these definitions for individual cells as well.
Namely,

\begin{definition}
	Given a configuration $ \sigma: \torus{m}{n} \to \bitset $ and a cell $ \ell \in \torus{m}{n} $,
	\begin{enumerate}
		\item we say that $ \ell $ is a \textsf{monochromatic cell} if $ \sigma(\ell) = 1 $ and there is a cell $ \ell' \in \Gamma_{=1}(\ell) $ satisfying $ \sigma(\ell') = 1 $.
		
		\item we say that $ \ell $ is a \textsf{chessboard cell} if every cell $ \ell' \in \Gamma_{=1}(\ell) $ satisfies $ \sigma(\ell') \ne \sigma(\ell) $ and $ \ell $ is not a \nameref*{def:moore} $ 1 $-\nameref*{def:isolated} cell.
	\end{enumerate}
\end{definition}

\begin{figure}
	\centerline{\mbox{\includegraphics[width=.83\textwidth]{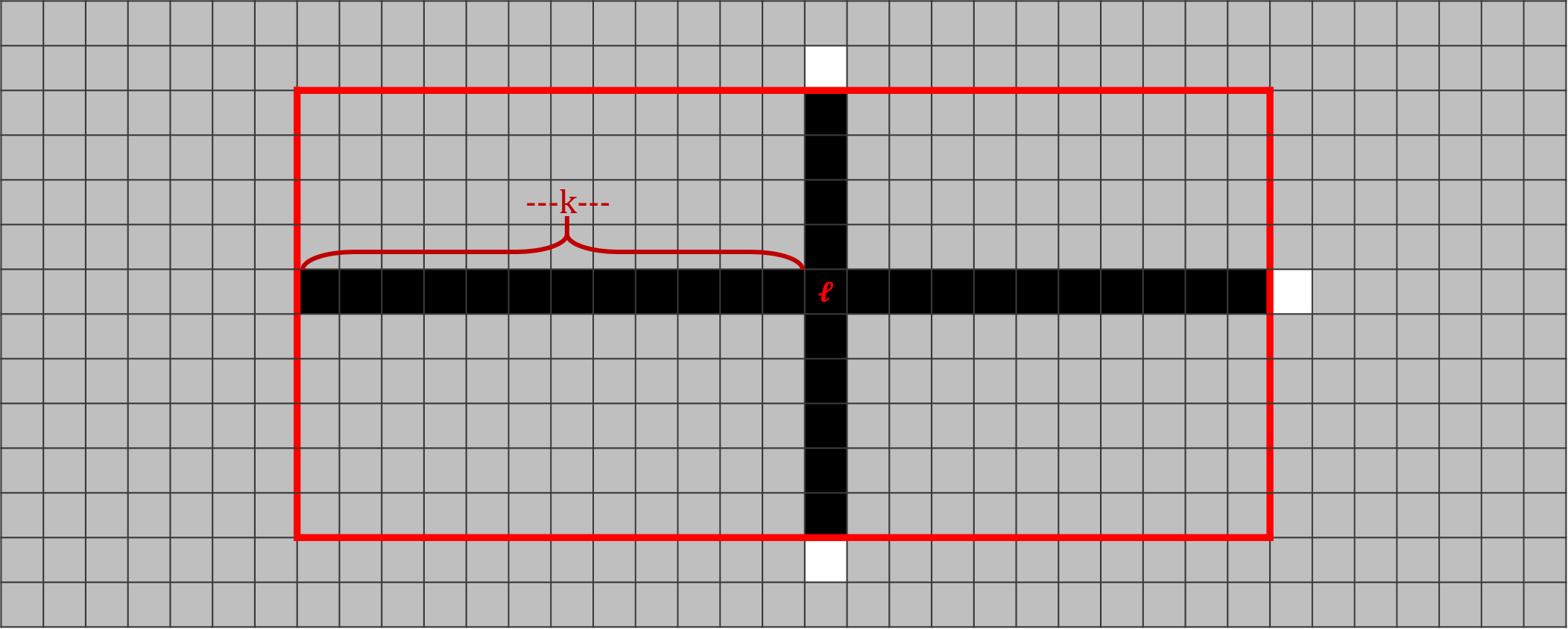}}}
	\caption{
		An example of the bounding box of the distance-$ k $ \nameref*{def:plus} of a monochromatic cell $ \ell $ for $ k=12 $.
		The dark cells in the figure correspond to state-$ 1 $ cells, the white ones correspond to state-$ 0 $ cells and the gray ones correspond to cells whose state is not irrelevant.
	}
	\label{fig:bounding_box}
\end{figure}

Now we define the \nameref*{def:plus} of a cell as well as its bounding box. For an illustration, see \Cref{fig:bounding_box}.

\begin{definition}[cross-shaped region]\label{def:plus}
	Given a configuration $ \sigma: \torus{m}{n} \to \bitset $ and a monochromatic (chessboard) cell $ (r,c) \in \torus{m}{n} $, we define the distance-$ k $ \textsf{monochromatic (chessboard) \nameref*{def:plus}} of the cell $ (r,c) $ as the maximal connected set of cells $ C \subseteq (\set{r} \times \cycnums{n}) \cup (\cycnums{m} \times \set{c}) $ such that $ (r,c) \in C $, every cell in $ C $ is a monochromatic (chessboard) cell and the distance between $ (r,c) $ and every cell in $ C $ is at most $ k $.
\end{definition}

\begin{definition}
	Given a configuration $ \sigma: \torus{m}{n} \to \bitset $ and a monochromatic/chessboard cell $ \ell \in \torus{m}{n} $, we denote by $ B_k^{\sigma}(\ell) $ the \textsf{bounding box}\footnote{
		Given a connected set of cells $ C \subseteq \torus{m}{n} $, let $ I = \set{i \in \cycnums{n} : (i,j) \in C} $ and $ J = \set{j \in \cycnums{n} : (i,j) \in C} $ (noting that since the set $ C $ is connected, the sets $ I $ and $ J $ are coordinate intervals), the \emph{bounding box} of the set $ C $ is the set $ I \times J $.
	} of the distance-$ k $ monochromatic/chessboard \nameref*{def:plus} of the cell $ \ell $.
	When the configuration in question is apparent from the context, we abbreviate $ B_k^{\sigma}(\ell) $ by $ B_k(\ell) $.
\end{definition}

\subsection{Interior and perimeter violations}
In addition to \nameref*{def:wraparound_violating_pair}s, we introduce two more types of violations: violations that are \textsf{interior} to the bounding box of a cell's \nameref*{def:plus} and violations that lie on the \emph{perimeter} of that bounding box.
In order to describe these two types of violations, we need to be able to precisely refer to the \textsf{parity} of the \textsf{distance} between cells in a set of cells that is free of \emph{odd} \nameref*{def:wraparound}s.
Formally,

\begin{definition}
	We say that a connected set of cells $ C \subseteq \torus{m}{n} $ contains an odd \nameref*{def:wraparound} if $ m $ is odd and there exists a cell $ (i,j) \in C $ for every $ j \in \cycnums{n} $, or if $ n $ is odd and there exists a cell $ (i,j) \in C $ for every $ i \in \cycnums{m} $.
	Otherwise, we say that $ C $ is \textsf{free} of odd \nameref*{def:wraparound}s.
\end{definition}

\begin{definition}
	Given a connected set of cells $ C \subseteq \torus{m}{n} $ that is free of odd \nameref*{def:wraparound}s and a pair of cells $ \ell_1, \ell_2 \in C $, we denote by $ \textsf{p}_C(\ell_1,\ell_2) $ the parity of the length of a path from $ \ell_1 $ to $ \ell_2 $ within $ C $.\footnote{Since the set $ C $ is free of odd \nameref*{def:wraparound}s, the lengths of all paths from $ \ell_1 $ to $ \ell_2 $ have the same parity.}
\end{definition}

\begin{figure}
	\centerline{\mbox{\includegraphics[width=.9\textwidth]{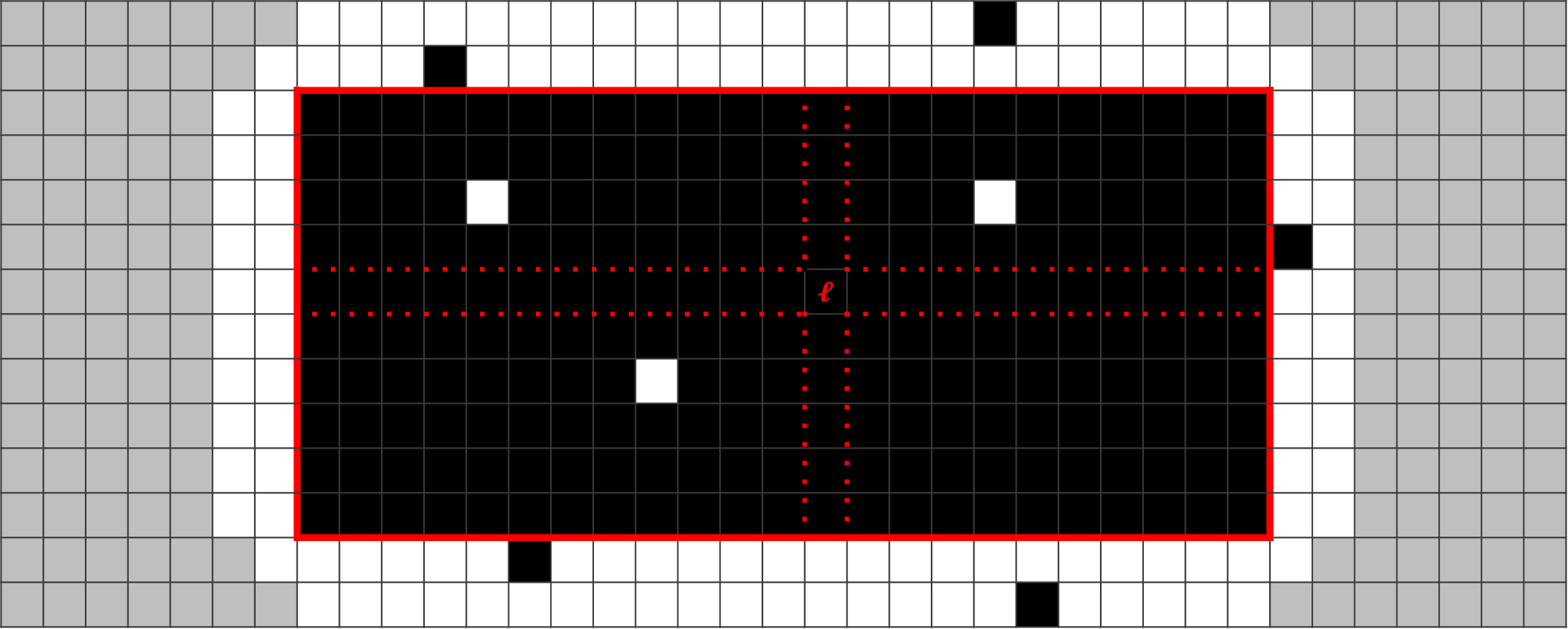}}}
	\caption{
		An example of the bounding box of the \nameref*{def:plus} of a monochromatic cell $ \ell $ with three interior violations and five perimeter violations.
		The dark and white cells correspond to state-$ 1 $ and state-$ 0 $ cells, respectively, and the gray cells are not relevant for the definition.
	}
	\label{fig:violations}
\end{figure}

We additionally define the distance-$ r $ neighborhood of a \emph{set of cells} $ C \subseteq \torus{m}{n} $ as the set of all cells with Manhattan distance at most $ r $ from \emph{any} cell in $ C $.
That is, for a set $ C \subseteq \torus{m}{n} $,
$ \Gamma_{\le r}(C) = \bigcup_{\ell \in C} \Gamma_{\le r}(\ell) $,
and, for $ r \ge 1 $,
$ \Gamma_{=r}(C) = \Gamma_{\le r}(C) \setminus \Gamma_{\le r-1}(C) $.
We can now introduce \textsf{interior} and \textsf{exterior} violations (see illustrations in \Cref{fig:violations}).

\begin{definition}
	Given a configuration $ \sigma: \torus{m}{n} \to \bitset $, let $ \ell \in \torus{m}{n} $ be a monochromatic or a chessboard cell in $ \sigma $.
	
	Given an integer $ k $, we say that a cell $ \ell' \in B_k(\ell) $ constitutes an \textsf{interior violation} with respect to $ B_k(\ell) $ if:
	\begin{itemize}
		\item $ \sigma(\ell') = 0 $ in the case where $ \ell $ is a monochromatic cell.
		\item $ \sigma(\ell') \ne \sigma(\ell) \oplus \textsf{p}_{B_k(\ell)}(\ell,\ell') $ in the case where $ \ell $ is a chessboard cell.
	\end{itemize}
	We say that a cell $ \ell' \in \Gamma_{=1}(B_k(\ell)) \cup \Gamma_{=2}(B_k(\ell)) $ constitutes a \textsf{perimeter violation} with respect to $ B_k(\ell) $ if:
	\begin{itemize}
		\item $ \sigma(\ell') = 1 $ in the case where $ \ell $ is a monochromatic cell.
		\item In the case where $ \ell $ is a chessboard cell, if $ \sigma(\ell') = 1 $ and one of the following holds:
		\begin{itemize}
			\item the cell $ \ell' $ belongs to $ \Gamma_{=1}(B_k(\ell)) $.
			\item the cell $ \ell' $ belongs to $ \Gamma_{=2}(B_k(\ell)) $ and $ \sigma(\ell) \oplus \textsf{p}_{\Gamma_{\le 2}(B_k(\ell))}(\ell,\ell') = 1 $.
			\item the cell $ \ell' $ belongs to $ \Gamma_{=2}(B_k(\ell)) $ and is monochromatic.
		\end{itemize}
	\end{itemize}
\end{definition}

\subsection{\nameref*{def:rectangulation}}
We now begin describing our testing algorithm.
A key element in the testing algorithm (as well as in the stabilization algorithm which we design to prove the testing algorithm's correctness) is the \nameref*{def:rectangulation} of a configuration.
The \nameref*{def:rectangulation} of a configuration is essentially a process where we partition the cells in $ \torus{m}{n} $ into  rectangles and set to $ 0 $ all the cells in the $ 3 $-\nameref*{def:boundary} of each rectangle.
The $ 3 $-\nameref*{def:boundary} of a rectangle is its $ 3 $-steps neighboring zone. Formally,

\begin{definition}[boundary]\label{def:boundary}
	Given a connected set of cells $ C \subseteq \torus{m}{n} $, we define the \textsf{$ 1 $-\nameref*{def:boundary}} of $ C $ as the set of cells $ \ell \in C $ that are adjacent to at least one cell outside of $ C $.
	For every $ k>1 $, we define the \textsf{$ k $-\nameref*{def:boundary}} of $ C $ as the set of cells $ \ell \in C $ that belong or are adjacent to the $ (k-1) $-\nameref*{def:boundary} of $ C $.
\end{definition}

\begin{definition}[Rectangulation]\label{def:rectangulation}
	Given a configuration $ \sigma: \torus{m}{n} \to \bitset $ and an integer $ k $, we define the \textsf{$ k $-\nameref*{def:rectangulation}} of the configuration $ \sigma $ as the configuration $ \sigma^{\#}: \torus{m}{n} $ obtained by executing \Cref{alg:rectangulation} (specified below) on $ \sigma $ as input.
\end{definition}

\begin{mdframed}[backgroundcolor=gray!20,linecolor=white]
\begin{algorithm}[H]\caption{Rectangulation algorithm}\label{alg:rectangulation}
	Input: a configuration $ \sigma: \torus{m}{n} \to \bitset $ and an integer $ k $.
	\begin{enumerate}
		\item Partition $ \torus{m}{n} $ into equal-size rectangles with $ k $ rows and $ k $ columns each.
		Let $ \mathcal{R} $ be the set of the rectangles in the partition.
		
		\item For each rectangle $ R \in \mathcal{R} $, set to $ 0 $ the state of all the cells belonging to the $ 1 $-\nameref*{def:boundary} of $ R $.
		
		\item For each rectangle $ R \in \mathcal{R} $, set to $ 0 $ every \nameref*{def:moore} $ 1 $-\nameref*{def:isolated} cell in the $ 3 $-\nameref*{def:boundary} of $ R $.
	\end{enumerate}
\end{algorithm}
\end{mdframed}

\begin{observation}
	Given query access to a configuration $ \sigma: \torus{m}{n} \to \bitset $, one has query access to the configuration $ \sigma^{\#} $, the \nameref*{def:rectangulation} of $ \sigma $.
\end{observation}

\subsection{The testing algorithm}
With the \nameref*{def:rectangulation} algorithm at hand, as well as having defined the three types of violations, we can finally specify our testing algorithm.

\begin{mdframed}[backgroundcolor=gray!20,linecolor=white]
\begin{algorithm}[H]\caption{Testing Algorithm for configuration stability under the 2D Threshold-2 rule}\label{alg:threshold2_adaptive_tester}
	\setcounter{step}{0}
	
	\begin{step}\label{step:chess_wraparound}
		\textbf{Detection of chessboard \nameref*{def:wraparound_violating_pair}s.}
		\begin{enumerate}
			\item Select, independently, uniformly at random $ \Theta(1/\eps) $ rows and $ \Theta(1/\eps) $ columns. 
			\item For each selected row/column, select, independently, uniformly at random $ \Theta(1/\eps) $ cells belonging to the row/column and query $ \Gamma_{\le 3}(\ell) $ for each selected cell $ \ell $.
			\item If the sample contains an unstable cell or a \nameref*{def:wraparound_violating_pair}, \textsf{Reject}.
		\end{enumerate}
	\end{step}
	
	\begin{step}\label{step:find_violations}
		\textbf{Detection of interior and perimeter violations.}
		\begin{enumerate}
			\item Set $ k = c_1/\eps $ for a constant $ c_1 $ we determine in the analysis.
			Let $ \sigma^{\#} $ be the $ k $-\nameref*{def:rectangulation} of the configuration $ \sigma $.
			For any monochromatic/chessboard cell $ \ell \in \torus{m}{n} $, we use the notation $ B_k(\ell) $ to refer to $ B_k^{\sigma^{\#}}(\ell) $.
			
			\item Select, independently, uniformly at random a sample $ S \subseteq \torus{m}{n} $ of $ \Theta(1/\eps) $ cells and query $ \Gamma_{\le 2}(\ell) $ for each cell $ \ell \in S $.
			
			\item\label{step:reject_unstable} If any cell $ \ell \in S $ is found to be unstable, \textsf{Reject}.
			
			\item For each monochromatic and each chessboard cell $ \ell \in S $ with respect to the configuration $ \sigma^{\#} $, make $ \Theta(1/\eps) $ queries to identify $ B_k(\ell) $, i.e., the bounding box of the cell's distance-$ k $ monochromatic/chessboard \nameref*{def:plus} in $ \sigma^{\#} $.
			
			\item For each monochromatic or chessboard cell $ \ell \in S $ with respect to the configuration $ \sigma^{\#} $, query the set $ \Gamma_{\le 3}(B_k(\ell)) \setminus B_k(\ell) $ in $ \sigma^{\#} $.
			If the set contains any perimeter violation with respect to $ B_k(\ell) $ in $ \sigma^{\#} $, \textsf{Reject}.
			
			\item For each monochromatic or chessboard cell $ \ell \in S $ with respect to the configuration $ \sigma^{\#} $, query all the cells in the $ 1 $-\nameref*{def:boundary} of $ B_k(\ell) $.
			If the set contains any interior violation with respect to $ B_k(\ell) $ in $ \sigma^{\#} $, \textsf{Reject}.
			
			\item For each monochromatic or chessboard cell $ \ell \in S $ with respect to the configuration $ \sigma^{\#} $, select, independently, uniformly at random a sample $ S_\ell \subseteq B_k(\ell) $ of $ \Theta(1/\eps) $ cells.
			If the sample contains any interior violation with respect to $ B_k(\ell) $ in $ \sigma^{\#} $, \textsf{Reject}.
			
			\item Otherwise, \textsf{Accept}.
		\end{enumerate}
	\end{step}
\end{algorithm}
\end{mdframed}

The query complexity of \Cref{alg:threshold2_adaptive_tester} is clearly $ \Theta(1/\eps^2) $.
Also, since \Cref{alg:threshold2_adaptive_tester} rejects a configuration only upon encountering a violation, and given that stable configurations are free of violations, \Cref{alg:threshold2_adaptive_tester} accepts every stable configuration with probability $ 1 $.
We still have to show, though, that \Cref{alg:threshold2_adaptive_tester} rejects every configuration that is $ \eps $-far from being stable with high constant probability.
That is, we have to show that if a configuration $ \sigma: \torus{m}{n} \to \bitset $ is $ \eps $-far from being stable, then it is either the case that \Cref{alg:threshold2_adaptive_tester} rejects $ \sigma $ with high constant probability or that $ \sigma $ can be \emph{stabilized} by modifying the states of at most $ \eps mn $ cells.
To show this, we describe a modification procedure that has that feature: \nameref*{alg:threshold2_adaptive_stabilization}.

\subsection{The stabilization algorithm}
We stress that the Stabilization Algorithm (\Cref{alg:threshold2_adaptive_stabilization}) is not part of the testing algorithm (\Cref{alg:threshold2_adaptive_tester}), but only serves the testing algorithm's analysis.
The heart of the Stabilization Algorithm lies in ``fixing'' the violations inside the bounding boxes of some \nameref*{def:plus}s, specifically the bounding boxes that contain ``relatively few'' violations.
We refer to such bounding boxes as \textsf{$ \alpha $-good}, where $ \alpha \in [0,1] $ (we apply the definition with $ \alpha = \Theta(\eps) $).

\begin{definition}[$ \alpha $-good bounding box]\label{def:good}
	Let $ \sigma: \torus{m}{n} \to \bitset $ be a configuration and $ \ell \in \torus{m}{n} $ be a monochromatic/chessboard cell in $ \sigma $.
	Given an integer $ k $ and a parameter $ 0 < \alpha \le 1 $, we say that $ B_k(\ell) $ is \textsf{$ \alpha $-good} with respect to $ \sigma $ if all of the following hold:
	\begin{enumerate}
		\item No cell $ \ell' \in \Gamma_{=1}(B_k(\ell)) \cup \Gamma_{=2}(B_k(\ell)) $ constitutes a perimeter violation with respect to $ B_k(\ell) $.
		\item No cell in the $ 1 $-\nameref*{def:boundary} of $ B_k(\ell) $ constitutes an internal violation with respect to $ B_k(\ell) $.
		\item At most an $ \alpha $-fraction of the cells $ \ell' \in B_k(\ell) $ constitutes an interior violation with respect to $ B_k(\ell) $.
	\end{enumerate}
\end{definition}

Before fixing violations within maximal $ \alpha $-good bounding boxes, we perform two steps.
In \Cref{step:fix_wraparounds} of the stabilization algorithm (see \Cref{alg:threshold2_adaptive_stabilization}), we fix chessboard $ \alpha $-\nameref*{def:wraparound_consistent} rows (or columns) and transform them into chessboard \nameref*{def:wraparound} rows (or columns).
This is done by invoking \Cref{alg:fix_wraparound} (which appears right after the stabilization algorithm).
When fixing such different rows (similarly, columns) and their distance-$ 2 $ von-Neumann Neighborhoods, there is no overlap in the modified cells, so we can think of performing these calls to \Cref{alg:fix_wraparound} ``in parallel''.

In \Cref{step:rectangulation}, we run the \nameref*{def:rectangulation} procedure (\Cref{alg:rectangulation}), \emph{except} that we do not modify cells belonging to the fixed \nameref*{def:wraparound} rows (columns).

Then, in \Cref{step:fix_good_boxes}, we fix the maximal bounding boxes, which are defined with respect to the \nameref*{def:rectangulation} (so that each such bounding box is entirely contained within one of the rectangles defined by the \nameref*{def:rectangulation}).
By the definition of maximal $ \alpha $-good bounding boxes, no two such bounding boxes can overlap (unless they are identical).
For bounding boxes that \emph{do not} intersect \nameref*{def:wraparound} rows (columns), the fixing step is straightforward (see \Cref{step:fix_good_boxes_no_intersection}).
For bounding boxes that \emph{do} intersect \nameref*{def:wraparound} rows (columns), the fixing is more intricate, and we defer the details to \Cref{sec:omitted_step}.

Finally, in \Cref{step:set_bad_cells_to_zero}, we set to $ 0 $ the states of all cells that do not belong to the fixed \nameref*{def:wraparound} rows (columns) or to maximal bounding boxes.

\begin{mdframed}[backgroundcolor=gray!20,linecolor=white]
\begin{algorithm}[H]\caption{The Threshold-2 Stabilization Algorithm}\label{alg:threshold2_adaptive_stabilization}
	\setcounter{step}{0}
	\newcounter{substep}[step]
	\renewcommand{\thesubstep}{\thestep.\arabic{substep}}
	Input: a configuration $ \sigma: \torus{m}{n} \to \bitset $.
	
	\begin{step}\label{step:fix_wraparounds}
		\textbf{Fixing chessboard $ \alpha $-\nameref*{def:wraparound_consistent} rows or columns.}
		\begin{enumerate}
			\item Set $ \alpha = \eps / c_2 $ for a constant $ c_2 $ whose value is set in the analysis.
			Let $ \mathcal{I} $ be the set of chessboard $ \alpha $-\nameref*{def:wraparound_consistent} rows and let $ \mathcal{J} $ be the set of chessboard $ \alpha $-\nameref*{def:wraparound_consistent} columns in $ \sigma $.
			
			\item If $ |\mathcal{I}| \ge |\mathcal{J}| $, invoke \Cref{alg:fix_wraparound} for all the rows in $ \mathcal{I} $, thereby transforming them into chessboard \nameref*{def:wraparound} rows, and let $ W \subseteq \torus{m}{n} $ be the set of cells belonging to the rows in $ \mathcal{I} $.
			Otherwise, modify \Cref{alg:fix_wraparound} to handle columns instead of rows, invoke it for all the columns in $ \mathcal{J} $, and let $ W $ be the set of cells belonging to the columns in $ \mathcal{J} $.
		\end{enumerate}
	\end{step}
	
	\begin{step}\label{step:rectangulation}
		\textbf{\nameref*{def:rectangulation}.}
		\begin{enumerate}
			\item Set $ k $ as in \Cref{alg:threshold2_adaptive_tester}.
			
			\item Execute the \nameref*{alg:rectangulation} (\Cref{alg:rectangulation}) on the original configuration $ \sigma $ with parameter $ k $, but without modifying the cells belonging to the set $ W $ (i.e., do not undo the fixing of the chessboard $ \alpha $-\nameref*{def:wraparound_consistent} rows/columns that was done in \Cref{step:fix_wraparounds}).
		\end{enumerate}
	\end{step}
	
	\begin{step}\label{step:fix_good_boxes}
		\textbf{Fixing the $ \alpha $-good bounding boxes.}
		\begin{enumerate}
			\item\refstepcounter{substep}\label{step:fix_good_boxes_no_intersection} Let $ \mathcal{B} $ be the set of maximal $ \alpha $-good bounding boxes of the distance-$ k $ \nameref*{def:plus}s of each monochromatic/chessboard cell $ \ell \in \torus{m}{n} $ in $ \sigma^{\#} $.
			
			\item\refstepcounter{substep} For each bounding box $ B \in \mathcal{B} $ such that $ B \cap \Gamma_{\le 2}(W) = \emptyset $, set the states of all the cells that constitute an interior violation with respect to $ B $, so as to eliminate these violations.
			
			\item\refstepcounter{substep}\label{step:wraparound_good_boxes_intersection}
			For each bounding box $ B \in \mathcal{B} $ such that $ B \cap \Gamma_{\le 2}(W) \ne \emptyset $,
			apply the procedure described in \Cref{sec:omitted_step}.
		\end{enumerate}
	\end{step}
	
	\begin{step}\label{step:set_bad_cells_to_zero}
		Set to $ 0 $ the state of each cell $ \ell $ that does not belong to any $ B \in \mathcal{B} $ or to $ W $.
	\end{step}
\end{algorithm}
\end{mdframed}

We next provide the algorithm invoked in \Cref{step:fix_wraparounds}, noting that a configuration $ \sigma: \torus{m}{n} \to \bitset $ can contain chessboard \nameref*{def:wraparound} rows (columns) only if $ n $ ($ m $) is even.

\begin{mdframed}[backgroundcolor=gray!20,linecolor=white]
\begin{algorithm}[H]\caption{Transform a row to chessboard \nameref*{def:wraparound}}\label{alg:fix_wraparound}
	\setcounter{step}{0}
	
	Input: a configuration $ \sigma: \torus{m}{n} \to \bitset $ s.t. $ n $ is even and a row $ r \in \cycnums{m} $ where it is either the case that the majority of the cells in row $ r $ are \textsf{even} chessboard \nameref*{def:wraparound_consistent} or that the majority of them are \textsf{odd} chessboard \nameref*{def:wraparound_consistent}.
	
	\begin{enumerate}
		\item If the majority of the cells in row $ r $ are even chessboard \nameref*{def:wraparound_consistent}, let $ \ell \in \torus{m}{n} $ be such a cell.
		Otherwise, let $ \ell \in \torus{m}{n} $ be an odd chessboard \nameref*{def:wraparound_consistent} cell.
		Set the state of every cell $ \ell' $ in row $ r $ to $ \textsf{p}_{\set{r} \times \cycnums{n}}(\ell, \ell') \oplus \sigma(\ell) $.
		
		\item Set the state of every cell in rows $ r-1 $ or $ r+1 $ to $ 0 $.
		
		\item\label{step:wraparoud_insulation} For every cell $ \ell' $ belonging to row $ r-2 $ or $ r+2 $, if $ \textsf{p}_{\set{r-2,\cdots,r+2} \times \cycnums{n}}(\ell, \ell') \oplus \sigma(\ell) = 1 $, set the state of $ \ell' $ to $ 0 $.
		
		\item Set the state of every monochromatic cell belonging to rows $ r-2 $ or $ r+2 $ to $ 0 $.
	\end{enumerate}
\end{algorithm}
\end{mdframed}

In \Cref{sec:stabilization_correctness}, we prove that for every configuration $ \sigma: \torus{m}{n} \to \bitset $, the configuration resulting from applying \Cref{alg:threshold2_adaptive_stabilization} on $ \sigma $ is indeed stable.
Then, in \Cref{sec:tester_correctness}, we prove that if the configuration $ \sigma $ is $ \eps $-far from being stable, then \Cref{alg:threshold2_adaptive_tester} rejects $ \sigma $ with high constant probability, thereby establishing \Cref{thm:testable}.

\subsection{Bounding boxes that intersect with wraparound rows}\label{sec:omitted_step}
Here we provide the details of \Cref{step:wraparound_good_boxes_intersection} of \Cref{alg:threshold2_adaptive_stabilization} (the stabilization algorithm).
Roughly speaking, in \Cref{step:wraparound_good_boxes_intersection}, which deals with bounding boxes $ B\in \mathcal{B} $ that intersect \nameref*{def:wraparound} rows (i.e., contain cells in $ W $) we fix the violation of cells in $ B \setminus \Gamma_{\le 2}(W) $. However, this is not quite precise.
First, we may need to modify cells in $ \Gamma_{=2}(W) $ so as to ensure distance constraints (when $ B $ is determined by a monochromatic cell) or to fix violations (when $ B $ is determined by a chessboard cell, and there are no conflicts with cells in $ W $).
Second, following the modification performed to cells in $ \Gamma_{\le 2}(W) $, we might be left with single rows in $ B $ that do not constitute legal rectangular \nameref*{def:component}s.
Details are given below.

\begin{mdframed}[backgroundcolor=gray!20,linecolor=white]
	\paragraph*{\Cref{step:wraparound_good_boxes_intersection}:}
	Let $ \sigma_{wa} $ be the configuration resulting from applying \Cref{step:fix_wraparounds} of the algorithm, and let $ \sigma_{wa}^\# $ be the configuration resulting from applying \Cref{step:fix_wraparounds} and \Cref{step:rectangulation}.
	In what follows we assume that rows of $ \sigma $, rather than columns, were modified.
	The case in which columns were modified is analogous.
	
	\noindent
	For each $ B \in \mathcal{B} $ such that $ B \cap \Gamma_{\leq 2}(W) \neq \emptyset $,  do:
	\begin{enumerate}	
		\item Let $ \ell_B $ be a cell for which $ B $ is the bounding box of its \nameref*{def:plus}.
		\item If $ \ell_B $ is a monochromatic cell in $ \sigma $, do:
		\begin{enumerate}
			\item \label{step:stable-mono-W2} Set the states of all the cells in $ B \cap \Gamma_{=2}(W) $ to $ 0 $.
			
			\item \label{step:stable-mono-B} Set the states of all the cells in $ B \setminus \Gamma_{\leq 2}(W) $ to $ 1 $ with the following exception.
			If $ B $ has width one, then every cell in $ B $ that is either at distance-$ 1 $ from two different cells in $ \Gamma_{=2}(W) $ or is at distance-$ 1 $ from one cell in $ \Gamma_{=2}(W) $ and a row in $\Gamma_{=1}(B) $ is set to  $ 0 $.
		\end{enumerate}
		\item Otherwise ($ \ell_B $ is a chessboard cell in $ \sigma $), do:
		\begin{enumerate}
			\item \label{step:stable-chess-W2} For each cell $ \ell' \in B \cap \Gamma_{=2}(W) $, do:
			\begin{enumerate}
				\item \label{step:stable-chess-W2-erase} If $\ell'$ is at distance-$2$ from two different cells in $ W $, or is at distance-$ 2 $ from one cell in $ W $ and at distance-$ 1 $ from a row in $ \Gamma_{=1}(B) $, then set $ \ell' $ to $ 0 $.
				\item \label{step:stable-chess-W2-mod} Else, let $\ell''$ be the single cell in $W$ that is at distance-$2$ from  $\ell'$, let $r$ be the row of $\ell'$, and let $v(\ell') = \textsf{pd}_{\set{r} \times \cycnums{n}}(\ell_B, \ell') \oplus \sigma(\ell_B)$.
				If $v(\ell') \neq \sigma_{wa}(\ell'')$, then set $\ell'$ to $v(\ell')$.
			\end{enumerate}
			\item \label{step:stable-chess-B} For each cell $\ell' \in B \setminus \Gamma_{\leq 2}(W)$, let $r$ be the row of $\ell'$ and set $\ell'$ to
			$v(\ell') = \textsf{pd}_{\set{r} \times \cycnums{n}}(\ell_B, \ell') \oplus \sigma(\ell_B)$ with the following exception. If all the cells in rows $r-1$ and $r+1$ belonging to $B$ were modified to $0$ in the previous item or in Step~1, or one of these rows was modified to $0$ and the other is at distance-$1$ from $ \Gamma_{=1}(B) $, then set $\ell'$ to $0$.
		\end{enumerate}
	\end{enumerate}
\end{mdframed}

\newpage
\subsection{Auxiliary claims and observations}\label{sec:tester_auxiliary}
In order to establish the correctness of the testing algorithm, we need to prove the correctness of the stabilization algorithm.
First, however, we state some observations and prove some claims that are necessary for the correctness of both the testing algorithm and the stabilization algorithm.
The correctness proof of the stabilization algorithm appears in \Cref{sec:stabilization_correctness} and the correctness of the testing algorithm appears in \Cref{sec:tester_correctness}.
The correctness of both rely on the structural result for the Threshold-$ 2 $ rule (\Cref{claim:limit_cycles_in_th2}) whose proof appears in \Cref{sec:threshold2_structure_proof}.

\begin{observation}\label{claim:moore_unstable}
	Given a configuration $ \sigma: \torus{m}{n} \to \bitset $ and a \nameref*{def:moore} \nameref*{def:isolated} $ 1 $-cell $ \ell \in \torus{m}{n} $ in $ \sigma $, if the cell $ \ell $ is not row or column chessboard \nameref*{def:wraparound_consistent}, then the \nameref*{def:moore} of $ \ell $ contains an unstable cell.
\end{observation}

\begin{observation}\label{claim:tester_accepts_stable}
	Given a configuration $ \sigma: \torus{m}{n} \to \bitset $ and a pair of cells $ \ell_1, \ell_2 \in \torus{m}{n} $ such that each of $ \ell_1 $ and $ \ell_2 $ is either a monochromatic cell or a chessboard cell, if both $ B_k(\ell_1) $ and $ B_k(\ell_2) $ are $ \alpha $-good with respect to the $ k $-\nameref*{def:rectangulation} of the configuration $ \sigma $ for some integer $ k $ and parameter $ 0 \le \alpha \le 1 $, then:
	\begin{enumerate}
		\item It is either the case that $ B_k(\ell_1) \cap B_k(\ell_2) = \emptyset $ or that $ B_k(\ell_1) \cap B_k(\ell_2) = B_k(\ell_1) $ or $ B_k(\ell_1) \cap B_k(\ell_2) = B_k(\ell_2) $.
		\item the distance between $ B_k(\ell_1) $ and $ B_k(\ell_2) $ is at least $ 2 $, and if any of $ \ell_1 $ or $ \ell_2 $ is a monochromatic cell, then the distance is at least $ 3 $.
	\end{enumerate}
\end{observation}

\begin{observation}\label{claim:wraparound_cost}
	Given a configuration $ \sigma: \torus{m}{n} \to \bitset $ and a parameter $ \alpha<1/2 $, if we apply \Cref{alg:fix_wraparound} to all the chessboard $ \alpha $-\nameref*{def:wraparound_consistent} rows of $ \sigma $, then the number of cells we modify is at most $ 5\alpha mn $.
\end{observation}

\begin{observation}
	Given a configuration $ \sigma: \torus{m}{n} \to \bitset $ and a parameter $ \alpha < 1/2 $, the result of applying \Cref{alg:fix_wraparound} to all the chessboard $ \alpha $-\nameref*{def:wraparound_consistent} rows in $ \sigma $ is a configuration in which all these rows are chessboard \nameref*{def:wraparound}s.
\end{observation}

\begin{observation}\label{claim:stabilization_algorithm_correctness}
	For every input configuration $ \sigma: \torus{m}{n} \to \bitset $, the output of \Cref{alg:threshold2_adaptive_stabilization} is a stable configuration.
\end{observation}

\begin{observation}\label{claim:bad_cells_dont_induce_good_rectangles}
	Let $ \sigma: \torus{m}{n} \to \bitset $ be a configuration and let $ \ell \in \torus{m}{n} $ be a cell whose state would be set to $ 0 $ in \Cref{step:set_bad_cells_to_zero} of \Cref{alg:threshold2_adaptive_stabilization}.
	If the cell $ \ell $ is stable in $ \sigma $, then $ B_k(\ell) $ is not $ \alpha $-good (for $ k $ and $ \alpha $ as set in \Cref{step:fix_good_boxes} of \Cref{alg:threshold2_adaptive_stabilization}).
\end{observation}

\begin{claim}\label{claim:B_intersects_W}
	Let $ B \in \mathcal{B} $ be such that $ B \cap \Gamma_{\leq 2}(W) \neq \emptyset $, and let $D(B,\Gamma_{\leq 2}(W))$ be the subset of cells in $B\cap \Gamma_{\leq 2}(W)$ that were modified in \Cref{step:fix_wraparounds} of \Cref{alg:threshold2_adaptive_stabilization}.
	That is,
	$D(B,\Gamma_{\leq 2}(W)) = \{\ell'\in B\cap \Gamma_{\leq 2}(W)\,:\,\sigma_{wa}(\ell') \neq \sigma(\ell)\}$.
	Then the number of cells in $B$ that were modified in \Cref{step:wraparound_good_boxes_intersection} is upper bounded by
	$2\cdot (\alpha\cdot |B| + |D(B,\Gamma_{\leq 2}(W))|)$.
\end{claim}
\begin{proof}
	In a nutshell, we prove the claim by showing that each cell that is modified in \Cref{step:wraparound_good_boxes_intersection} is either a violating cell with respect to its bounding box $B$, or it can be mapped to either a violating cell in $B$ or a cell that is modified in \Cref{step:fix_wraparounds}, where the mapping is at most $ 2 $-to-$ 1 $.
	
	We break the analysis into three cases (and within them, into sub-cases).
	
	\paragraph*{$\ell_B$ is a monochromatic cell in $\sigma$ and the width of $B$ is greater than $1$.}
	Let $V(B)$ be the subset of violating cells in $B$ with respect to $\sigma$, that is, whose state is $0$ in $\sigma$.
	First observe that each cell in $B \setminus \Gamma_{\leq 2}(W)$ that is modified in
	\Cref{step:stable-mono-B} (from $0$ to $1$) belongs to $V(B)$.
	We next turn to consider cells in $B \cap \Gamma_{=2}(W)$ that were modified in \Cref{step:stable-mono-W2} (from $1$ to $0$).
	
	Recall that in the execution of \Cref{alg:fix_wraparound} (invoked in \Cref{step:chess_wraparound} of \Cref{alg:threshold2_adaptive_stabilization}), every cell in $B \cap \Gamma_{=2}(W)$ that is monochromatic in $\sigma$,  is set to $0$.
	Each such cell has state-$1$ and at least one neighboring cell with state-$1$.
	Therefore, each cell $\ell'$ in $B \cap \Gamma_{=2}(W)$ that is modified in \Cref{step:stable-mono-W2} from $1$ to $0$, must have a neighboring cell $\ell''$ in $B$ that belongs to its row and has state $0$ in $\sigma$.
	That is, $\ell'' \in V(B)$.
	Observing that each cell in $V(B)\cap \Gamma_{=2}(W)$ is not modified and is adjacent to at most two cells in its row (in particular, that belong to $B \cap \Gamma_{=2}(W)$ and are modified), we get the following.
	The total number of cells in $B$ that are modified in the execution of Items~\ref{step:stable-mono-B} and~\ref{step:stable-mono-W2} is at most $2|V(B)|$.
	Since $B \in \mathcal{B}$, we have that $|V(B)| \leq \alpha |B|$, and so the number of these cells is upper bounded by $2\alpha |B|$.

	\paragraph*{$\ell_B$ is a monochromatic cell in $\sigma$ and the width of $B$ is $1$.}
	The analysis of this case is the same as for the previous case, except for the special sub-case addressed at the end of \Cref{step:stable-mono-B}.
	That is, we need to account also for cells $\ell'$ belonging to $B\cap \Gamma_{=3}(W)$ that were modified from $1$ to $0$ because they are at distance-$1$ from two different cells in $\Gamma_{=2}(W)$ (i.e., cells that belong to two different rows: one above and one below) or at distance-$1$ from one such cell in $\Gamma_{=2}(W)$ and from one cell in $ \Gamma_{=1}(B) $.
	
	For each such cell $\ell'$, let $\ell''$ be a cell in $\Gamma_{=1}(W)$ at distance-$2$ from $\ell'$.
	If $\ell''$ has state-$0$ in $\sigma$, then it belongs to $V(B)$ (and it is not modified in any step).
	Otherwise, it was modified from state-$1$ to state-$0$ in the transformation to $\sigma_{wa}$, so that it belongs to $D(B,\Gamma_{\leq 2}(W))$.
	Each such cell $\ell''$ is at distance-$2$ from at most two cells $\ell' \in B\cap \Gamma_{=3}(W)$.
	Hence (combining the argument from the previous case with the one above concerning cells $\ell' \in B\cap \Gamma_{=3}(W)$), in this case the total number of cells in $B$ that are modified in
	Items~\ref{step:stable-mono-B} and~\ref{step:stable-mono-W2} is at most $2|V(B)| + 2|D(B,\Gamma_{\leq 2}(W))| \leq \alpha|B| + 2|D(B,\Gamma_{\leq 2}(W))|$.
	
	\paragraph*{$\ell_B$ is a chessboard cell in $\sigma$.}
	Here too let  $V(B)$ be the subset of violating cells in $B$ with respect to $\sigma$ and $\ell_B$.
	Consider first the cells in $B \cap \Gamma_{=2}(W)$ that are modified in \Cref{step:stable-chess-W2-erase}.
	For each such cell $\ell'$, let $\ell'$ be a cell in $W$ at distance-$2$ from $\ell'$.
	We claim that either $\ell'$ belongs to $V(B)$ or $\ell''$ belongs to $V(B) \cup D(B,\Gamma_{\leq 2}(W))$.
	To verify this, first observe that if $\sigma_{wa}^{\#}(\ell')=0$, then no modification is performed.
	Hence, $\sigma_{wa}^{\#}(\ell')=1$, implying that $\sigma(\ell')=1$ and that $\sigma_{wa}(\ell'')=0$.
	Since $v(\ell') = v(\ell'')$, if $v(\ell')=0$, then $\ell'\in V(B)$, and if $v(\ell')=1$, then either $\ell''\in V(B)$ (when $\sigma(\ell'')=\sigma_{wa}(\ell'')$) or $\ell'' \in D(B,\Gamma_{\leq 2}(W))$.
	
	Next consider the cells in $B \cap \Gamma_{=2}(W)$ that are modified in \Cref{step:stable-chess-W2-mod}.
	The state of each such cell $\ell'$ is set to $v(\ell')$ (after verifying that $v(\ell') \neq \sigma_{wa}(\ell'')$).
	Therefore, the state of $\ell'$ is modified only if it belongs to $V(B)$.
	
	Finally we turn to cells in $B \setminus \Gamma_{\leq 2}(W)$ that are modified in \Cref{step:stable-chess-B}.
	Similarly to \Cref{step:stable-mono-B}, one case in which a cell $\ell' \in B\setminus \Gamma_{\leq 2}(W)$ is modified in this step is when $\sigma(\ell') \neq v(\ell')$, implying that $\ell' \in V(B)$.
	Another case is in the exception clause in \Cref{step:stable-chess-B}.
	Namely, where $\ell'\in \Gamma_{=3}(W)$ is modified from state-1 to state-0, even though it is not in $V(B)$, because all the cells in rows $r-1$ and $r+1$ belonging to $B$ were modified to $0$ in the previous item or in Step~1, or one of these rows was modified to $0$ and the other is on the outer distance-$1$ boundary of $B$.
	For each such cell $\ell'$ let $\ell''$ be a cell at distance-$2$ from it in $\Gamma_{=1}(W)$ (so that $\ell''$ has state $0$).
	Here too $v(\ell') = v(\ell'')$.
	If $v(\ell') = 0$, then $\ell' \in V(B)$.
	Otherwise ($v(\ell')=1$), then either $\ell'' \in V(B)$ (because $\sigma(\ell'')=0 \neq v(\ell'')$) or $\ell'' \in D(B,\Gamma_{\leq 2}(W))$ (because $\sigma(\ell'')=1 \neq \sigma_{wa}(\ell'')$).
	
	We note that in the analysis of \Cref{step:stable-chess-W2-erase}, modified cells in $B \cap \Gamma_{=2}(W)$ that do not belong to $V(B)$ were mapped to (``charge'') cells in $W$ that belong to $V(B) \cup D(B,\Gamma_{\leq 2}(W))$, while in the analysis of \Cref{step:stable-chess-B}, cells in $B\setminus \Gamma_{\leq 2}(W)$ that do not belong to $V(B)$ were mapped to (``charge'') cells in $\Gamma_{=1}(W)$ that belong to $V(B) \cup D(B,\Gamma_{\leq 2}(W))$.
	
	We have thus established that in this case (where $\ell(B)$ is a chessboard cell in $\sigma$), for every modified cell $\ell'\in B$, either $\ell' \in V(B)$, or it can be mapped to a cell $\ell''$ in $W\cup \Gamma_{=1}(W)$ such that $\ell''\in V(B)\cup D(B,\Gamma_{\leq 2}(W))$, and to any such cell $\ell''$, at most two cells in $B$ are mapped.
	Hence the total number of modified cells is at most $2|V(B)| + 2|D(B,\Gamma_{\leq 2}(W))| \leq \alpha|B| + 2|D(B,\Gamma_{\leq 2}(W))|$.
\end{proof}

\subsection{Correctness of the stabilization algorithm (\Cref*{alg:threshold2_adaptive_stabilization})}\label{sec:stabilization_correctness}
\begin{claim}
	For every configuration $ \sigma: \torus{m}{n} \to \bitset $, the configuration resulting from applying \Cref{alg:threshold2_adaptive_stabilization} on $ \sigma $ is a stable configuration.
\end{claim}
\begin{proof}
	In order to prove that after running the stabilization algorithm on any configuration $\sigma$ we obtain a stable configuration (with respect to the Threshold-2 rule), we establish two claims.
	The first is  that every state-$1$ cell in the final configuration belongs either to a rectangular $1$-\nameref*{def:mono_component} or to a rectangular \nameref*{def:chess_component}.
	The second is that every two of these \nameref*{def:component}s are at distance at least $2$, and if at least one of them is a $1$-\nameref*{def:mono_component}, then the distance is at least $3$.
	By \Cref{claim:limit_cycles_in_th2}, this suffices to imply that the configuration is stable.
	
	For the first claim, consider any state-$1$ cell $\ell$ in the final configuration, which we denote by $\sigma_f$.
	The cell $\ell$ either belongs to $W$ or to some maximal $\alpha$-good bounding box $B$ (in $\mathcal{B}$).
	
	Consider first the case that $\ell\in W$.
	By the description of \Cref{alg:fix_wraparound}, which we invoke in \Cref{step:fix_wraparounds} of the stabilization algorithm, after \Cref{step:fix_wraparounds} is completed, all state-$1$ cells in $W$ belong to chessboard wraparound rectangles (where either all are rows or all are columns).
	Furthermore, in all following steps, cells in $\Gamma_{\leq 1}(W)$ are not further modified, and cells in $\Gamma_{=2}(W)$ may be modified only from state-$1$ to  state-$0$, with one exception.
	The exception is when they belong to a maximal $\alpha$-good bounding box of the chessboard type, when for every such cell that obtains state-$1$, the cell at distance-$2$ from it in  $W$ has state-$0$.
	This ensures that the \nameref*{def:wraparound} rows (columns) with cells in $W$ are indeed \nameref*{def:chess_component}s (i.e., without violating maximality).
	
	Next we turn to the case that $\ell \in B$ for some $B\in \mathcal{B}$ (\Cref{step:fix_good_boxes}).
	If cells in $B$ are modified in \Cref{step:fix_good_boxes_no_intersection}, then, clearly, if $B$ is determined by a monochromatic cell, then (following this step), all state-$1$ cells in $B$ belong to a rectangular state-$1$ \nameref*{def:mono_component}, and if it is determined by a chessboard cell, then all state-$1$ cells in $B$ belong to a rectangular \nameref*{def:chess_component}.
	If the cells in $B$ are modified in \Cref{step:wraparound_good_boxes_intersection}, then we need to consider two sub-cases, as discussed next.
	For the sake of simplicity, we assume that the cells in $W$ belong to \nameref*{def:wraparound} rows (the argument is easily adapted to the case in which they are \nameref*{def:wraparound} columns).
	
	The first sub-case is when $\ell$ belongs to a bounding box $B\in \mathcal{B}$ that is determined by a monochromatic cell.
	In this case, by \Cref{step:stable-mono-B} in \Cref{step:wraparound_good_boxes_intersection} of the stabilization algorithm, $\ell$  belongs to a rectangular state-$1$ \nameref*{def:mono_component} in $\sigma_f$.
	In particular, in the special case that $B$ has width one, it is ensured that every state-$1$ cell in $B$ has at least one such adjacent cell.
	
	The second sub-case is when $B$ is determined by a chessboard cell.
	By \Cref{step:stable-chess-W2} in \Cref{step:wraparound_good_boxes_intersection}, each cell $\ell\in B$ that has state-$1$ in $\sigma_f$ must belong to a rectangle of height at least two (whose columns are all columns of $B$).
	
	It remains to verify that the distance constraints are obeyed in the final configuration $\sigma_f$.
	First consider any pair of \nameref*{def:component}s $C$ and $C'$ in $\sigma_f$ (which by the above are rectangular), such that at least one of them, say $C$, is a state-$1$ \nameref*{def:mono_component}.
	By the description of the algorithm, the cells of $C$ must belong to some maximal bounding box $B \in \mathcal{B}$.
	Let the rows of $C$ be in the interval $[a,a']$ and the columns in $[b,b']$ (where the latter are also the columns of $B$).
	Consider any cell $(x,y)$ on the boundary of $C$.
	If $y=b$, then, since $B$ is $ \alpha $-good with respect to $\sigma^\#$, we have that $\sigma^\#(x,b-1)=\sigma^\#(x,b-2)=0$, and the state of both these cells is $0$ in $\sigma_f$ as well.
	An analogous statement holds for the case that $y=b'$.
	If $x= a$, then we have that $\sigma_f(a-1,y)=\sigma_f(a-2,y)=0$ either due to $B$ being $ \alpha $-good with respect to $\sigma^\#$, (in the case that $a$ is the first row in $B$) or due to the combination of \Cref{step:wraparoud_insulation} in \Cref{alg:fix_wraparound} and \Cref{step:stable-mono-W2} in \Cref{step:wraparound_good_boxes_intersection} of the stabilization algorithm.
	An analogous statement holds for the case that $x=a'$.
	Hence, the distance between $C$ and any other \nameref*{def:component} $C'$ is necessarily at least $ 3 $ in $\sigma_f$.
	
	It remains to consider pairs of \nameref*{def:component}s $C$ and $C'$ that are both \nameref*{def:chess_component}s in $\sigma_f$.
	If at least one of them is a \nameref*{def:wraparound} \nameref*{def:component}, then the distance of at least $ 2 $ between them is ensured by \Cref{step:wraparoud_insulation} in \Cref{alg:fix_wraparound}.
	If they both correspond to (parts of) bounding boxes in $\mathcal{B}$, then the distance requirement follows from the bounding box being $ \alpha $-good.
\end{proof}

\subsection{Correctness of the testing algorithm (\Cref*{alg:threshold2_adaptive_tester})}\label{sec:tester_correctness}
As stated in \Cref{sec:testing_alg}, \Cref{alg:threshold2_adaptive_tester} accepts every stable configuration with probability $ 1 $.
This is because stable configurations are free of violations, and the algorithm is designed to reject only when encountering a violation.
Hence, to complete the proof of the algorithm's correctness, it is left to show that the algorithm rejects every configuration that is $ \eps $-far from being stable with high constant probability.

\begin{lemma}
	If a configuration $ \sigma: \torus{m}{n} $ is $ \eps $-far from being stable, then \Cref{alg:threshold2_adaptive_tester} rejects $ \sigma $ with probability at least $ 2/3 $.
\end{lemma}
\begin{proof}
	Let $ \sigma: \torus{m}{n} \to \bitset $ be any configuration.
	We prove that it is either the case that \Cref{alg:threshold2_adaptive_tester}, our testing algorithm, rejects $ \sigma $ with probability at least $ 2/3 $, or that the total number of cells modified by \Cref{alg:threshold2_adaptive_stabilization} (\nameref*{alg:threshold2_adaptive_stabilization}) is smaller than $ \eps mn $.
	As the latter case implies that the configuration $ \sigma $ is $ \eps $-close to being stable, this will establish the claim.
	
	Recall that $ k = c_1/\eps $ (as set in \Cref{alg:threshold2_adaptive_tester}) and $ \alpha = \eps / c_2 $ (as set in \Cref{alg:threshold2_adaptive_stabilization}).
	We first claim that the total number of cells modified in Steps \ref{step:fix_wraparounds}-\ref{step:fix_good_boxes} of \Cref{alg:threshold2_adaptive_stabilization} is at most $ (\frac{12}{c_1} + \frac{17}{c_2}) \eps mn $:
	\begin{enumerate}
		\item By \Cref{claim:wraparound_cost}, the total number of cells we modify in \Cref{step:fix_wraparounds} is at most $ 5\alpha mn \le 5\epsilon mn / c_2 $.
		
		\item In \Cref{step:rectangulation}, the \nameref*{def:rectangulation} step, the total number of rectangles in the partition is $ mn / k^2 $ and the number of cells we modify in each rectangle is at most $ 3 $ times its perimeter, i.e., $ 12k $.
		Thus, the total number of modifications made in \Cref{step:rectangulation} is at most $ 12k \cdot mn/k^2 \le 12 \eps mn / c_1 $.
		
		\item In \Cref{step:fix_good_boxes}, since each rectangle $ B \in \mathcal{B} $ is $ \alpha $-good, the number of cells we modify in each rectangle $ B $ is at most $ \alpha |B| $.
		By \Cref{claim:B_intersects_W}, for every $ B \in \mathcal{B} $ whose cells are modified in \Cref{step:wraparound_good_boxes_intersection} of the stabilization algorithm, the number of cells in $ B $ that were modified in this step is upper bounded by
		$ 2\cdot (\alpha\cdot |B| + |D(B,\Gamma_{\leq 2}(W))|) $, where $ D(B,\Gamma_{\leq 2}(W)) $ is the subset of cells in $ B\cap \Gamma_{\leq 2}(W) $ that were modified in \Cref{step:chess_wraparound} of the algorithm.
		As noted above, the total number of cells modified in \Cref{step:fix_wraparounds} of the algorithm is at most $ 5\eps mn/c_2 $.
		Since the bounding boxes in $ \mathcal{B} $ are disjoint, the total number of cells modified in \Cref{step:fix_good_boxes} is at most $ 2\alpha m n + 10\eps mn/c_2 \leq 12\eps m n/c_2 $.
	\end{enumerate}
	
	We now turn to bounding the number of cells modified in \Cref{step:set_bad_cells_to_zero} of \Cref{alg:threshold2_adaptive_stabilization}.
	Let $ X \subseteq \torus{m}{n} $ be the set of cells that neither belong to $ W $ nor to any set $ B \in \mathcal{B} $.
	Each cell $ \ell $ that is set to $ 0 $ in \Cref{step:set_bad_cells_to_zero} belongs to the set $ X $, and, therefore, is either a \nameref*{def:moore} $ 1 $-\nameref*{def:isolated} cell or a monochromatic/chessboard cell for which $ B_k^{\sigma^{\#}}(\ell) $ is not $ \alpha $-good.
	We define three subsets of the set $ X $, whose union equals $ X $, such that for each of these subset, either the number of cells in the subset is sufficiently small or \Cref{alg:threshold2_adaptive_tester} rejects the configuration $ \sigma $ with probability at least $ 2/3 $.
	Let $ c_3 $ be a constant that we set later, where, in what follows, we assume that all the constants implicit in the $ \Theta(\cdot) $ notation throughout \Cref{alg:threshold2_adaptive_tester} are sufficiently large with respect to the constant $ c_3 $.
	
	\begin{enumerate}
		\item Let $ X_1 \subseteq X $ be the set of chessboard/monochromatic cells $ \ell \in \torus{m}{n} $ for which $ B_k^{\sigma^{\#}}(\ell) $ is not $ \alpha $-good.
		It is either the case that $ |X_1| \le \eps mn / c_3 $ or that \Cref{alg:threshold2_adaptive_tester} rejects the configuration $ \sigma $ with probability at least $ 2/3 $ in \Cref{step:find_violations}.
		
		\item Let $ X_2 \subseteq X $ be the set of \nameref*{def:moore} $ 1 $-\nameref*{def:isolated} cell in $ X $ that are not \nameref*{def:wraparound_consistent}.
		By \Cref{claim:moore_unstable}, every \nameref*{def:moore} $ 1 $-\nameref*{def:isolated} cell that is not \nameref*{def:wraparound_consistent} has an unstable cell in its \nameref*{def:moore}.
		If there are more than $ \eps mn / (9c_3) $ unstable cells, then \Cref{alg:threshold2_adaptive_tester} rejects the configuration $ \sigma $ in \Cref{step:reject_unstable} of \Cref{step:find_violations} with probability at least $ 2/3 $.
		Thus, it is either the case that there are at most $ |X_2| \le 9 \eps mn / (9c_3) \le \eps mn / c_3 $, or that \Cref{alg:threshold2_adaptive_tester} rejects the configuration $ \sigma $ with probability at least $ 2/3 $.
		
		\item Let $ X_3 \subseteq X $ be the set of \nameref*{def:moore} $ 1 $-\nameref*{def:isolated} cells that are \nameref*{def:wraparound_consistent} (outside of $ W $ and of any $ B \in \mathcal{B} $).
		Without loss of generality, we assume that $ |\mathcal{I}| \ge |\mathcal{J}| $.
		If there are fewer than $ \eps m / (2c_3) $ rows with at least $ \epsilon n / (2c_3) $ \nameref*{def:wraparound_consistent} \nameref*{def:moore} $ 1 $-\nameref*{def:isolated} cells outside of $ W $ and of any $ B \in \mathcal{B} $, then $ |X_3| \le \eps mn / (2c_3) + \eps mn / (2c_3) = \eps mn / c_3 $ (at most $ n $ such cells in at most $ \eps m / (2c_3) $ rows plus at most $ \eps n / (2c_3) $ such cells in at most $ m $ rows).
		Otherwise, \Cref{alg:threshold2_adaptive_tester} detects a \nameref*{def:wraparound_violating_pair} with probability at least $ 2/3 $ in \Cref{step:chess_wraparound}.
	\end{enumerate}
	
	Hence, if we set the constants $ c_1, c_2 $ and $ c_3 $ such that $ \frac{12}{c_1} + \frac{17}{c_2} + \frac{1}{c_3} < 1 $, then it is either the case that \Cref{alg:threshold2_adaptive_tester} rejects the configuration $ \sigma $ with probability at least $ 2/3 $ or that \Cref{alg:threshold2_adaptive_stabilization} modifies fewer than $ \eps mn $ cells.
\end{proof}
\section{Proof of the Threshold-2 structural result}\label{sec:threshold2_structure_proof}
In this section we prove \Cref{claim:limit_cycles_in_th2}, the structural characterization of the stable configurations of the Threshold-2 rule on the two-dimensional torus.
The proofs of the two directions of the claim are in \Cref{sec:threshold2_structure_proof_directions}, where we use some auxiliary claims and observations stated and proved in \Cref{sec:threshold2_structure_auxiliary}.

First, we restate definitions of \nameref*{def:mono_component}s and of \nameref*{def:chess_component}s in terms of an explicit list of requirements that we can refer back to in our proofs.

\begin{definition}[monochromatic component]\label{def:mono_component}
	Given a configuration $ \sigma : \torus{m}{n} \to \bitset $ and a value $ \beta \in \bitset $, we say that a set $ C \in \torus{m}{n} $ is a \textsf{state-$ \beta $ \nameref*{def:mono_component}} if all of the following hold.
	\begin{enumerate}
		\item $ C $ is connected.
		\item Every cell $ \ell \in C $ satisfies $ \sigma(\ell) = \beta $.
		\item For every cell $ \ell \in C $, there is at least one other cell in $ C $ that is adjacent to $ \ell $.
		Equivalently, given the other conditions, $ |C| \ge 2 $.
		\item $ C $ is not strictly contained in another set that satisfies the previous conditions.
	\end{enumerate}
\end{definition}

\begin{definition}[chessboard component]\label{def:chess_component}
	Given a configuration $ \sigma : \torus{m}{n} \to \bitset $, we say that a set $ C \in \torus{m}{n} $ is a \textsf{\nameref*{def:chess_component}} if all of the following hold.
	\begin{enumerate}
		\item\label{req:chess_component_connected} $ C $ is connected.
		\item\label{req:chess_component_chessy} For every pair of adjacent cells $ \ell, \ell' \in C $, $ \sigma(\ell) \ne \sigma(\ell') $.
		\item\label{req:chess_component_no_lugs} For every cell $ \ell \in C $, there are at least two other cells in $ C $ that are adjacent to $ \ell $.
		\item\label{req:chess_component_maximality} $ C $ is not strictly contained in another set that satisfies the previous conditions.
	\end{enumerate}
\end{definition}

\begin{definition}[component]\label{def:component}
	Given a configuration $ \sigma : \torus{n} \to \bitset $ and a set of cells $ C \subseteq \torus{n} $, if $ C $ is either a state-1 \nameref*{def:mono_component} or a \nameref*{def:chess_component} in $ \sigma $, we say that $ C $ is a \textsf{\nameref*{def:component} in $ \sigma $}.
\end{definition}

\subsection{Auxiliary claims and observations}\label{sec:threshold2_structure_auxiliary}
\begin{observation}\label{claim:neighborhoods_of_adjacent_cells_intersect}
	If two cells $ \ell, \ell' \in \torus{m}{n} $ are adjacent to each other, then all but one of the four cells in the set $ \Gamma_{=1}(\ell') $ are adjacent to a cell belonging to the set $ \Gamma_{=1}(\ell) \setminus \set{\ell'} $.
\end{observation}

\begin{observation}\label{claim:adjacent_1s_are_final_in_thr2}
	Given a configuration $ \sigma : \torus{m}{n} \to \bitset $, if $ \ell, \ell' \in \torus{m}{n} $ are two adjacent cells and $ \sigma(\ell) = \sigma(\ell') = 1 $, then $ \sigma_t(\ell) = \sigma_t(\ell') = 1 $ for every integer $ t $, where $ \sigma_t = \thr_2^t(\sigma) $.
	In particular, both $ \ell $ and $ \ell' $ are fixed in the configuration $ \sigma $.
\end{observation}

\begin{observation}\label{claim:connected_sets_of_1s_are_final}
	Given a configuration $ \sigma : \torus{m}{n} \to \bitset $, if $ C \subseteq \torus{m}{n} $ is a state-1 \nameref*{def:mono_component}, then for every cell $ \ell \in C $, $ \sigma_t(\ell) = 1 $ for every integer $ t $, where $ \sigma_t = \thr_2^t(\sigma) $.
	In particular, every cell $ \ell \in C $ is fixed in the configuration $ \sigma $.
\end{observation}

\begin{claim}\label{claim:cell_adjacent_to_either_zero_or_one_cell_in_a_rectangle}
	If a cell $ \ell \in \torus{m}{n} $ is adjacent to a rectangle $ R \subseteq \torus{m}{n} $ and $ R $ is not an almost-wraparound rectangle, then $ \ell $ is adjacent to exactly one cell in $ R $.
\end{claim}
\begin{proof}
	Let $ I \in \cycnums{m} $ and $ J \in \cycnums{n} $ be the two coordinate intervals defining the rectangle $ R $.
	That is, $ R = I \times J $.
	Let $ (i,j) \in \torus{m}{n} \setminus R $ be a cell that is adjacent to $ R $, and suppose by way of contradiction that there are at least two distinct cells $ (i',j'), (i'',j'') \in R \cap \Gamma_{\le 1}(\ell) $.
	
	Suppose first that $ i'=i'' $.
	In this case, $ i=i'=i'' $, and since $ i',i'' \in I $, it must also be the case that $ i \in I $.
	Since the cell $ (i,j) $ is adjacent to both $ (i,j') $ and $ (i,j'') $, we can assume without loss of generality that $ j'=j-1 $ and $ j''=j+1 $.
	That is, both $ (i,j-1) $ and $ (i,j+1) $ belong to the rectangle $ R $.
	This implies that either $ [j-1,j+1] \subseteq J $ or that $ [j+1,j-1] \subseteq J $.
	If the former holds, then $ j \in [j-1,j+1] \subseteq J $.
	If the latter holds, then it is either the case that $ J = \cycnums{n} $, in which case $ j \in J $ as well, or that $ J = [j+1,j-1] $, in which case $ R $ is an almost-wraparound rectangle, which we have assumed is not the case.
	Therefore, under the assumption that $ i'=i'' $, both $ i \in I $ and $ j \in J $, and therefore $ (i,j) \in R $, and we reach a contradiction.
	
	A similar argument can be carried out for the case that $ j'=j'' $.
	Suppose, then, that both $ i' \ne i'' $ and $ j' \ne j'' $.
	
	Since the cell $ (i,j) $ is adjacent to both $ (i',j') $ and $ (i'',j'') $, one of these two cells must be in row $ i $ and the other in column $ j $.
	Hence, we can assume without loss of generality that $ i'=i $ and $ j''=j $.
	Therefore, since $ (i,j') \in R $, it must be the case that $ i \in I $, and since $ (i'',j) \in R $, it must also be the case that $ j \in J $.
	Thus, $ (i,j) \in R $, and, again, we reach a contradiction.
\end{proof}

\begin{claim}\label{claim:cell_adjacent_to_at_least_two_cells_in_a_non_rectangle}
	If a set of cells $ C \subseteq \torus{m}{n} $ is connected but not a rectangle, then there exists a cell $ \ell \in \torus{m}{n} \setminus C $ s.t. $ |C \cap \Gamma_{\le 1}(\ell)| \ge 2 $.
	Moreover, there exists such a cell that belongs to the bounding box of the set $ C $.
\end{claim}
\begin{proof}
	Let $ C \subseteq \torus{m}{n} $ be a connected set of cells and assume $ C $ is not a rectangle.
	
	Given any rectangle $ R \subseteq C $, where $ R = I \times J $ for a pair of coordinate intervals $ I \in \cycnums{m}, J \subseteq \cycnums{n} $, we say that $ R $ can be \emph{vertically extended within the set $ C $} if $ I = [a,b] \subsetneq \cycnums{m} $, and it is either the case that $ (a-1,j) \in C $ for every $ j \in J $, or $ (b+1,j) \in C $ for every $ j \in J $.
	Similarly, we say that $ R $ can be \emph{horizontally extended within the set $ C $} if $ J = [c,d] \subsetneq \cycnums{n} $ and it is either the case that $ (i,c-1) \in C $ for every $ i \in I $, or $ (i,d+1) \in C $ for every $ i \in I $.
	Note that the set of cells resulting from \emph{extending a rectangle within the set $ C $} is a rectangle and a subset of $ C $.
	
	We claim that there exists a rectangle $ R \subseteq C $ that cannot be extended within $ C $.
	Suppose such a rectangle does not exist.
	In this case, we can take any cell in $ C $, and since the set consisting of that cell is a rectangle within $ C $, we can repeatedly \emph{extend it}, and since every time we extend a rectangle, we strictly increase the number of cells in the resulting set, it must be impossible to do this indefinitely, as the set $ C $ is finite.
	
	Let $ R \subseteq C $, then, be a rectangle that cannot be extended within $ C $.
	Since the set $ C $ is connected and $ C \ne R $ (because $ C $ is not a rectangle and $ R $ is), there must be a cell in $ C \setminus R $ that is adjacent to $ R $.
	
	Without loss of generality, we assume that the cell's first coordinate is $ a-1 $ where $ R = [a,b] \times J $.
	Since the rectangle $ R $ cannot be extended within $ C $, there has to be at least one cell that does not belong to $ C $ whose first coordinate is also $ a-1 $ and the second coordinate belongs to $ J $.
	Let $ (a-1,j) $ and $ (a-1,j+1) $ be a pair of cells such that one belongs to $ C $, the other does not belong to $ C $, and $ [j,j+1] \subseteq J $.
	
	The rectangle $ \set{(a,j), (a,j+1), (a-1,j), (a-1,j+1)} $ is a $ 2 \times 2 $ rectangle that intersects with $ C $ on exactly 3 cells.
	This implies that the one cell in this $ 2 \times 2 $ rectangle that does not belong to $ C $ is adjacent to at least 2 cells in $ C $, and the first part of the \namecref{claim:cell_adjacent_to_at_least_two_cells_in_a_non_rectangle} follows.
	
	Moreover, that cell shares its two coordinates with two other cells in the $ 2 \times 2 $ rectangle.
	As these two other cells belong to $ C $, this implies that the cell also belongs to the bounding box of $ C $.
\end{proof}

\begin{observation}\label{claim:chess_components_closure}
	Given a configuration $ \sigma : \torus{m}{n} \to \bitset $, let $ \ell \in \torus{m}{n} $ be a cell satisfying $ \sigma(\ell) = \beta $ for some $ \beta \in \bitset $ and let $ \ell', \ell'' \in \torus{m}{n} $ be a pair of distinct cells in $ \Gamma_{=1}(\ell) $ where each of $ \ell' $ and $ \ell'' $ belongs to some \nameref*{def:chess_component} in $ \sigma $ (not necessarily the same \nameref*{def:chess_component}).
	If $ \sigma(\ell') = \sigma(\ell'') \ne \beta $, then all the three cells $ \ell $, $ \ell' $ and $ \ell'' $ belong to the same \nameref*{def:chess_component} in $ \sigma $.
\end{observation}

\begin{observation}\label{claim:chess_component_closure_two_cells}
	Given a configuration $ \sigma : \torus{m}{n} \to \bitset $ and a \nameref*{def:chess_component} $ C \subseteq \torus{m}{n} $ in $ \sigma $, let $ \ell_1, \ell_2 \in \torus{m}{n} $ be a pair of adjacent cells satisfying $ \sigma(\ell_1) \ne \sigma(\ell_2) $.
	If there exists a pair of cells $ \ell_1' \in C \cap \Gamma_{\le 1}(\ell_1) $ and $ \ell_2' \in C \cap \Gamma_{\le 1}(\ell_2) $ satisfying $ \sigma(\ell_1) \ne \sigma(\ell_1') $, and $ \sigma(\ell_2) \ne \sigma(\ell_2') $, then $ \ell_1, \ell_2 \in C $.
\end{observation}

\begin{claim}\label{claim:X_and_Y_disjoint_in_temporally_periodic_configurations}
	If a configuration $ \sigma : \torus{m}{n} \to \bitset $ is stable, then no cell belongs both to a \nameref*{def:chess_component} and to a state-1 \nameref*{def:mono_component}.
\end{claim}
\begin{proof}
	Let $ \sigma'=\thr_2(\sigma) $ and $ \sigma'' = \thr_2(\sigma') $.
	Suppose, contrary to the claim, that there exists a cell $ \ell \in \torus{m}{n} $ that belongs both to a \nameref*{def:chess_component} and to a state-1 \nameref*{def:mono_component}, and denote the \nameref*{def:chess_component} by $ C $.
	Let $ \ell' \in C $ be a cell in $ \Gamma_{=1}(\ell) $.
	By requirement~(\ref{req:chess_component_chessy}) of \Cref{def:chess_component}, it must be the case that $ \sigma(\ell') = 0 $ and that there are at least two cells in $ \Gamma_{\le 1}(\ell') $ whose states in $ \sigma $ is 1.
	Hence, $ \sigma'(\ell') = 1 $.
	Since $ \ell $ belongs to a state-1 \nameref*{def:mono_component}, by \Cref{claim:connected_sets_of_1s_are_final}, $ \sigma'(\ell) = 1 $ as well.
	Thus, $ \sigma''(\ell') = 1 $, in contradiction to $ \sigma $ being stable.
\end{proof}

\subsection{The Proof of \Cref*{claim:limit_cycles_in_th2}}\label{sec:threshold2_structure_proof_directions}
\begin{proof}[Proof of first direction of \Cref{claim:limit_cycles_in_th2}]
	Let $ \sigma : \torus{m}{n} \to \bitset $ be a configuration that satisfies requirements (\ref{item:all_components_are_rectangles})-(\ref{item:rectangles_not_too_close}) for $ X $, $ Y $ and $ Z $ as defined in the \namecref{claim:limit_cycles_in_th2} statement.
	We show that $ \sigma $ is stable.
	Let $ \sigma'=\thr_2(\sigma) $ and $ \sigma'' = \thr_2(\sigma') $.
	
	We show that every cell $ \ell \in \torus{m}{n} $ satisfies $ \sigma'(\ell) \ne \sigma(\ell) $ if and only if $ \ell \in Y $.
	Assuming that this is true, we are done since we will be able to apply the same argument for the configuration $ \sigma' $ and get that every cell outside of $ Y $ is fixed and every cell in $ Y $ is toggling.
	Let $ \ell \in \torus{m}{n} $ be any cell.
	
	Suppose first that $ \ell \in Y $.
	That is, $ \ell \in R $ where $ R $ is a \nameref*{def:chess_component} in $ \sigma $ and also a rectangle.
	If $ \sigma(\ell) = 0 $, then, by the requirements specified in the definition of \nameref*{def:chess_component}s, the cell $ \ell $ is adjacent to at least two cells in $ R $ and the state of both of these cells must be $ 1 $ in $ \sigma $.
	Hence, $ \sigma'(\ell) = 1 $.
	If, on the other hand, $ \sigma(\ell) = 1 $, then all the cells which are adjacent to $ \ell $ inside $ R $ are in a state of $ 0 $ in $ \sigma $.
	In addition, since the distance between $ R $ and any other set $ C \in \mathcal{X} \cup \mathcal{Y} $ is at least 2, every cell outside of $ R $ that is adjacent to $ \ell $ must belong to $ Z $, and therefore that cell's state in $ \sigma $ is also $ 0 $.
	In other words, the state of each cell that is adjacent to $ \ell $, whether inside or outside of $ R $, must be $ 0 $ in $ \sigma $, implying that $ \sigma'(\ell) = 0 $.
	Thus, in any case, if $ \ell \in Y $, then $ \sigma'(\ell) \ne \sigma(\ell) $.
	
	Suppose now that $ \ell \notin Y $.
	
	If $ \ell \in X $, which means $ \ell \in R $, $ R $ being a state-1 \nameref*{def:mono_component} in $ \sigma $, then the cell $ \ell $ is adjacent to at least one cell in $ R $, and the state of that cell in $ \sigma $ must also be $ 1 $.
	Therefore, in this case, there are at least two 1s in $ \sigma[\Gamma_{\le 1}(\ell)] $, and, consequently, $ \sigma'(\ell) = 1 = \sigma(\ell) $ as well.
	
	Suppose, then, that $ \ell \in Z $.
	This means that $ \sigma(\ell) = 0 $, and so we need to show that $ \sigma'(\ell) = 0 $ as well.
	Suppose by way of contradiction that $ \sigma'(\ell) = 1 $.
	That is, there are at least two cells $ \ell', \ell'' \in \Gamma_{\le 1}(\ell) $ where $ \sigma(\ell') = \sigma(\ell'') = 1 $.
	Hence, neither $ \ell' $ nor $ \ell'' $ belong to $ Z $.
	If one of $ \ell' $ or $ \ell'' $ belongs to some $ R \in \mathcal{X} $, then, as $ R $ is not an almost-wraparound rectangle, by \Cref{claim:cell_adjacent_to_either_zero_or_one_cell_in_a_rectangle}, the other cell cannot belong to $ R $, and since the distance between $ R $ and any other set $ C \in \mathcal{X} \cup \mathcal{Y} $ is at least 3, the other cell cannot belong to any other set $ C \in \mathcal{X} \cup \mathcal{Y} $ besides $ R $.
	Hence, both $ \ell' $ and $ \ell'' $ belong to two, not necessarily distinct, sets $ R', R'' \in \mathcal{Y} $.
	However, since $ \sigma(\ell) = 0 $ and $ \sigma(\ell') = \sigma(\ell'') = 1 $, this implies that, by \Cref{claim:chess_components_closure}, $ \ell \in Y $, and we reach a contradiction.
\end{proof}

\begin{proof}[Proof of second direction of \Cref{claim:limit_cycles_in_th2}]
	Let $ \sigma : \torus{m}{n} \to \bitset $ be a stable configuration.
	Let $ \sigma' = \thr_2(\sigma) $ and $ \sigma'' = \thr_2(\sigma') $.
	We show that requirements (\ref{item:all_components_are_rectangles})-(\ref{item:rectangles_not_too_close}) hold for $ \sigma $.
	
	We begin by proving requirement (\ref{item:all_components_are_rectangles}), which is that every set $ C \in \mathcal{X} \cup \mathcal{Y} $ is a rectangle and no set $ C \in \mathcal{X} $ is an almost-wraparound rectangle.
	
	\paragraph*{Requirement~(\ref{item:all_components_are_rectangles}):}
	Suppose that, contrary to the claim, requirement (\ref{item:all_components_are_rectangles}) does not hold, and let $ C \in \mathcal{X} \cup \mathcal{Y} $ be a set that violates requirement (\ref{item:all_components_are_rectangles}).
	
	We first claim that if $ C \in \mathcal{X} $, then there exists a cell $ \ell \notin C $ that is adjacent to at least two distinct cells $ \ell', \ell'' \in C $.
	If the set $ C $ is not a rectangle, this follows directly from \Cref{claim:cell_adjacent_to_at_least_two_cells_in_a_non_rectangle}.
	If the set $ C $ is an almost-wraparound rectangle, then we can assume without loss of generality that $ C = I \times J $, where $ I \subseteq \cycnums{m} $ and $ J = \cycnums{n} \setminus \set{j} $ for some $ j \in \cycnums{n} $.
	For any $ i \in I $, then, the cell $ (i,j) $ does not belong to $ C $, and each of the two cells $ (i,j-1) $ and $ (i,j+1) $, which are adjacent to $ (i,j) $, does belong to $ C $.
	Since $ C \in \mathcal{X} $, by definition, the set $ C $ is a state-1 \nameref*{def:mono_component} in $ \sigma $, and since $ \ell $ is adjacent to $ C $, it must be the case that $ \sigma(\ell) = 0 $.
	\Cref{claim:connected_sets_of_1s_are_final} implies that $ \ell' $ and $ \ell'' $ are fixed in $ \sigma $, and since $ \sigma(\ell') = \sigma(\ell'') = 1 $, it must be the case that $ \sigma'(\ell) = \sigma'(\ell') = \sigma'(\ell'') = 1 $ and, by \Cref{claim:adjacent_1s_are_final_in_thr2}, the cell $ \ell $ is fixed in $ \sigma' $, implying that $ \sigma''(\ell) = 1 \ne \sigma(\ell) $, in contradiction to $ \sigma $ being stable.
	Hence, if $ C \in \mathcal{X} $, then $ C $ must be a rectangle, but not an almost-wraparound rectangle, as specified by requirement (\ref{item:all_components_are_rectangles}).
	
	Suppose next that $ C \in \mathcal{Y} $.
	If $ C $ is not a rectangle, then, again by \Cref{claim:cell_adjacent_to_at_least_two_cells_in_a_non_rectangle}, there is a cell $ \ell \notin C $ that is adjacent to at least two cells belonging to $ C $.
	Denote these two cells by $ \ell_1 $ and $ \ell_2 $.
	
	Suppose first that $ \sigma(\ell) = 1 $.
	If $ \sigma(\ell_1) = \sigma(\ell_2) = 0 $, then \Cref{claim:chess_components_closure} implies that $ \ell \in C $, and we reach a contradiction.
	Otherwise, it is not the case that $ \sigma(\ell_1) = \sigma(\ell_2) = 0 $, so we assume without loss of generality that $ \sigma(\ell_1) = 1 $.
	Since $ \ell_1 \in C $ and $ C $ is a \nameref*{def:chess_component} in $ \sigma $, the cell $ \ell_1 $ must be adjacent to at least two other cells in $ C $ whose states in $ \sigma $ are both 0.
	Let $ \ell_1' $ be one of these two cells.
	Since $ \sigma(\ell_1') = 0 $, and $ \ell_1' $ belongs to a \nameref*{def:chess_component}, there must be at least two cells in $ C $ that are adjacent to $ \ell_1' $ whose states are 1 in $ \sigma $.
	Hence, $ \sigma'(\ell_1') = 1 $.
	Since the cells $ \ell $ and $ \ell_1 $ are adjacent to each other and satisfy $ \sigma(\ell) = \sigma(\ell_1) = 1 $, it must be the case that $ \sigma'(\ell_1) = 1 $.
	Likewise, as $ \ell_1 $ and $ \ell_1' $ are adjacent to each other and satisfy $ \sigma'(\ell_1) = \sigma'(\ell_1') = 1 $, it must be the case that $ \sigma''(\ell_1') = 1 \ne \sigma(\ell_1') $, in contradiction to $ \sigma $ being stable.
	
	That is, under the assumption that $ C \in \mathcal{Y} $ and that $ C $ is not a rectangle, we ruled out the possibility that $ \sigma(\ell) = 1 $.
	It must, then, be the case that $ \sigma(\ell) = 0 $.
	If $ \sigma(\ell_1) = \sigma(\ell_2) = 1 $, then \Cref{claim:chess_components_closure} implies that $ \ell \in C $, and we reach a contradiction.
	If one of $ \sigma(\ell_1) $ and $ \sigma(\ell_2) $ equals 1 and the other equals 0, then we also reach a contradiction, because, since both $ \ell_1 $ and $ \ell_2 $ are adjacent to $ \ell $, the Manhattan distance between $ \ell_1 $ and $ \ell_2 $ is 2, so they must have the same state.
	Therefore, $ \sigma(\ell_1) = \sigma(\ell_2) = 0 $.
	This implies that $ \sigma'(\ell_1) = \sigma'(\ell_2) = 1 $, because each of $ \ell_1 $ and $ \ell_2 $ must be adjacent to at least two cells whose state in $ \sigma $ is 1, as they belong to a \nameref*{def:chess_component} in $ \sigma $.
	This means, however, that $ \sigma''(\ell) = 1 \ne \sigma(\ell) $, and we reach a contradiction.
	
	This concludes the proof that requirement (\ref{item:all_components_are_rectangles}) holds.
	We now turn to requirement (\ref{item:everything_outside_components_zero}), which is that every cell $ \ell \in Z $ satisfies $ \sigma(\ell) = 0 $.
	
	\paragraph*{Requirement~(\ref{item:everything_outside_components_zero}):}
	Suppose that, contrary to the claim, there exists a cell $ \ell \in Z $ satisfying $ \sigma(\ell) = 1 $.
	
	We claim that this implies that $ \Gamma_{\le 1}(\ell) \cap Z $ contains at least two cells, which we denote by $ \ell_{+1} $ and $ \ell_{-1} $, where $ \sigma(\ell_{+1}) = \sigma(\ell_{-1}) = 0 $, and that there are two cells, $ \ell_{+2} \in \Gamma_{\le 1}(\ell_{+1}) \cap Z \setminus \set{\ell} $ and $ \ell_{-2} \in \Gamma_{\le 1}(\ell_{-1}) \cap Z \setminus \set{\ell} $, where $ \sigma(\ell_{+2}) = \sigma(\ell_{-2}) = 1 $.
	
	If the cell $ \ell $ is adjacent to any cell $ \ell' $ where $ \sigma(\ell') = 1 $, then $ \ell $ belongs to a state-1 \nameref*{def:mono_component} and we reach a contradiction to the assumption that $ \ell \in Z $.
	Hence, every cell $ \ell' \in \Gamma_{=1}(\ell) $ satisfies $ \sigma(\ell') = 0 $.
	This implies that $ \sigma'(\ell) = 0 $, and since $ \sigma $ is stable, it must be the case that $ \sigma''(\ell) = 1 $.
	Hence, at least two of the cells $ \ell' \in \Gamma_{=1}(\ell) $ must satisfy $ \sigma'(\ell') = 1 $.
	We denote these two cells by $ \ell_{+1} $ and $ \ell_{-1} $ and show that both $ \ell_{+1} $ and $ \ell_{-1} $ belong to $ Z $.
	
	If $ \ell_{+1} $ belongs to a \nameref*{def:chess_component} $ C $, then there must exist two cells $ \ell_{+1}', \ell_{+1}'' \in \Gamma_{=1}(\ell_{+1}) \setminus \set{\ell} $ where $ \ell_{+1}', \ell_{+1}'' \in C $ and $ \sigma(\ell_{+1}') = \sigma(\ell_{+1}'') = 1 $.
	By \Cref{claim:neighborhoods_of_adjacent_cells_intersect}, at least one of these two cells, $ \ell_{+1}' $ or $ \ell_{+1}'' $, must be adjacent to a cell $ \ell_{+1}^* \in \Gamma_{\le 1}(\ell) \setminus \set{\ell_{+1}} $.
	Without loss of generality, assume that that cell is $ \ell_{+1}' $.
	Since the cell $ \ell_{+1}^* $ is adjacent to the cell $ \ell $, it must be the case that $ \sigma(\ell_{+1}^*) = 0 $.
	This implies, however, by \Cref{claim:chess_component_closure_two_cells}, that the pair of adjacent cells $ \ell $ and $ \ell_{+1}^* $ must belong to $ C $, but $ \ell $ belonging to a \nameref*{def:chess_component} is in contradiction to the assumption that $ \ell \in Z $.
	The same argument holds for the cell $ \ell_{-1} $.
	That is, both $ \ell_{+1} $ and $ \ell_{-1} $ belong to $ Z $.
	
	We now show that the cell $ \ell_{+1} $ is adjacent to a cell $ \ell_{+2} \in Z \setminus \set{\ell} $ satisfying $ \sigma(\ell_{+2}) = 1 $.
	Since $ \sigma(\ell_{+1}) = 0 $ and $ \sigma'(\ell_{+1}) = 1 $, the cell $ \ell_{+1} $ must be adjacent to at least one cell $ \ell_{+2} \in \torus{m}{n} \setminus \set{\ell} $ satisfying $ \sigma(\ell_{+2}) = 1 $.
	We claim that $ \ell_{+2} $ must belong to $ Z $.
	If $ \ell_{+2} $ belongs to a state-1 \nameref*{def:mono_component} in $ \sigma $, then by \Cref{claim:connected_sets_of_1s_are_final}, $ \sigma'(\ell_{+2}) = 1 $, and since $ \sigma'(\ell_{+1}) = 1 $ too, this would imply that $ \sigma''(\ell_{+1}) = 1 \ne \sigma(\ell_{+1}) $, and we reach a contradiction to $ \sigma $ being stable.
	Suppose, then, towards a contradiction, that the cell $ \ell_{+2} $ belongs to a \nameref*{def:chess_component} $ C \subseteq \torus{m}{n} $.
	Since $ \sigma(\ell_{+2}) = 1 $ and, since the cell $ \ell_{+2} $ cannot belong to a state-1 \nameref*{def:mono_component}, all the cells $ \ell^* \in \Gamma_{=1}(\ell_{+2}) $ satisfy $ \sigma(\ell^*) = 0 $.
	Hence, if there is a cell $ \ell^* \ne \ell_{+1} $ in $ \Gamma_{\le 1}(\ell_{+2}) \cap \Gamma_{\le 1}(\ell) $, then the set of cells $ \set{\ell, \ell_{+1}, \ell_{+2}, \ell^*} $ is $ 2 \times 2 $ rectangle in which every pair of adjacent cells in the set differ in their states in $ \sigma $.
	This implies, however, that all the cells in the set belong to $ C $, and we reach a contradiction to $ \ell \in Z $.
	We can therefore assume that $ \Gamma_{\le 1}(\ell_{+2}) \cap \Gamma_{\le 1}(\ell) = \set{\ell_{+1}} $.
	
	Since the cells $ \ell_{+2} $ and $ \ell_{+1} $ are adjacent, there must exist a cell $ \ell' \in \Gamma_{\le 1}(\ell_{+2}) \cap C $ such that the set $ \Gamma_{\le 1}(\ell_{+1}) \cap \Gamma_{\le 1}(\ell') $ contains a cell $ \ell'' $ which is distinct from $ \ell_{+2} $.
	Since $ \sigma'(\ell_{+1}) = 1 $ (because $ \ell_{+1} $ is adjacent to $ \ell $ and to $ \ell_{+2} $ and $ \sigma(\ell) = \sigma(\ell_{+2}) = 1 $) and $ \sigma'(\ell') = 1 $ (because $ \sigma(\ell') = 0 $ and $ \ell' \in C $, which means that $ \ell' $ is adjacent to at least two cells whose states in $ \sigma $ are 1), it must be the case that $ \sigma''(\ell'') = 1 $.
	Hence, since $ \sigma $ is stable, $ \sigma(\ell'') = 1 $.
	Thus, the set $ \set{\ell_{+1}, \ell_{+2}, \ell', \ell''} $ is $ 2 \times 2 $ rectangle in which every pair of adjacent cells in the set differ in their states in $ \sigma $, which implies, again, that all the cells in that set belong to $ C $, and we reach a contradiction to $ \ell_{+1} \in Z $.
	
	We have shown, then, that the cell $ \ell_{+1} $ is adjacent to a cell $ \ell_{+2} \in Z \setminus \set{\ell} $ satisfying $ \sigma(\ell_{+2}) = 1 $, and, by the same argument, the cell $ \ell_{-1} $ is adjacent to a cell $ \ell_{-2} \in Z \setminus \set{\ell} $ satisfying $ \sigma(\ell_{-2}) = 1 $.
	
	Noting that each of the two cells $ \ell_{+2} $ and $ \ell_{-2} $ satisfies the conditions we used to derive the claim for the cell $ \ell $, denoting $ \ell_0 = \ell $, we we can repeatedly apply the argument above and establish that there exists an infinite sequence of cells $ \set{\ell_{k}}_{k=-\infty}^{+\infty} $ with the following property:
	for every $ k \in \mathbb{Z} $, the two cells $ \ell_{2k+1} $ and $ \ell_{2k-1} $ belong to $ \Gamma_{\le 1}(\ell_{2k}) \cap Z $ and satisfy $ \sigma(\ell_{2k+1}) = \sigma(\ell_{2k-1}) = 0 $.
	Additionally, $ \ell_{2k+2} \in \Gamma_{\le 1}(\ell_{2k+1}) \cap Z \setminus \set{\ell_{2k}} $ and $ \ell_{2k-2} \in \Gamma_{\le 1}(\ell_{2k-1}) \cap Z \setminus \set{\ell} $, where $ \sigma(\ell_{2k+2}) = \sigma(\ell_{2k-2}) = 1 $.
	In particular, each pair of consecutive cells in the sequence are adjacent and differ in their state in $ \sigma $.
	Since the number of distinct cells in the sequence must be finite, the sequence contains a cycle, and since each pair of adjacent cells in the cycle differ in their states, the cycle must be contained in a \nameref*{def:chess_component} in contradiction to the assumption that for every $ k \in \mathbb{Z} $, $ \ell_k \in Z $.
	
	This concludes the proof that requirement (\ref{item:everything_outside_components_zero}) holds.
	We now turn to proving requirement~(\ref{item:rectangles_not_too_close}), which is that for every pair of distinct sets $ R,R' \in \mathcal{X} \cup \mathcal{Y} $, the distance between $ R $ and $ R' $ is at least 2, and if either $ R $ or $ R' $ belong to $ \mathcal{X} $, then the distance is at least 3.
	
	\paragraph*{Requirement~(\ref{item:rectangles_not_too_close}):}
	Let $ R \subseteq \torus{m}{n} $ be any set in $ \mathcal{X} \cup \mathcal{Y} $ and let $ R' \subseteq \torus{m}{n} $ be a set in $ \mathcal{X} \cup \mathcal{Y} $ whose distance from $ R $ is the smallest such that $ R' \ne R $.
	Suppose first that $ R \in \mathcal{X} $ and assume by way of contradiction that the distance between $ R $ and $ R' $ is smaller than 3.
	If the distance between $ R $ and $ R' $ is 0, it means that $ R \cap R' \ne \emptyset $.
	This, however, is impossible, because if $ R' \in \mathcal{X} $, then both $ R $ and $ R' $ are \nameref*{def:mono_component}s, so $ R \cap R' \ne \emptyset $ implies $ R = R' $, and if $ R' \in \mathcal{Y} $, we reach a contradiction to \Cref{claim:X_and_Y_disjoint_in_temporally_periodic_configurations} by which $ X \cap Y = \emptyset $.
	Hence, the distance between $ R $ and $ R' $ must be greater that $ 0 $.
	
	Let $ \ell \in R $ and $ \ell' \in R' $ be two (distinct) cells with minimum Manhattan distance.
	In particular, that distance is either 1 or 2.
	Suppose first that $ \sigma(\ell') = 1 $.
	In this case, the Manhattan distance between $ \ell $ and $ \ell' $ cannot be 1, because this would imply that $ \ell' \in R $, and this is impossible, because if $ R' \in \mathcal{X} $, then $ R=R' $ in contradiction to the assumption that $ R \ne R' $, and $ R' \in \mathcal{Y} $ contradicts \Cref{claim:X_and_Y_disjoint_in_temporally_periodic_configurations} by which $ X \cap Y = \emptyset $.

	Hence, the Manhattan distance between $ R $ and $ R' $ must be 2, which means that there is a cell $ \ell^* $ in $ \Gamma_{\le 1}(\ell) \cap \Gamma_{\le 1}(\ell') $.
	The cell $ \ell^* $ must belong to $ Z $ (otherwise, the set in $ \mathcal{X} \cup \mathcal{Y} $ that contains the cell $ \ell^* $ would be closer to $ R $ than the set $ R' $), which means, by requirement~(\ref{item:everything_outside_components_zero}), that $ \sigma(\ell^*)=0 $.
	Since $ \sigma(\ell) = \sigma(\ell') = 1 $, it must be the case that $ \sigma'(\ell^*) = 1 $, and since $ \ell $ belongs to a state-1 \nameref*{def:mono_component} in $ \sigma $, by \Cref{claim:connected_sets_of_1s_are_final}, $ \sigma'(\ell) = 1 $.
	Hence, $ \sigma''(\ell^*) = 1 \ne \sigma(\ell^*) $ in contradiction to $ \sigma $ being stable.
	Thus, it cannot be the case that $ \sigma(\ell') = 1 $.
	
	If $ \sigma(\ell') = 0 $, then $ R' $ must be a \nameref*{def:chess_component}, and this means that the cell $ \ell' $ is adjacent to at least two cells whose state in $ \sigma $ is 1, which implies that $ \sigma'(\ell') = 1 $.
	The same argument can be equivalently carried out for the configuration $ \sigma' $ instead of for $ \sigma $.
	
	It is left to show that if both $ R $ and $ R' $ are \nameref*{def:chess_component}s, then the distance between $ R $ and $ R' $ is at least 2.
	If the distance is 0, then there is a cell $ \ell^* \in R \cap R' $, which by \Cref{def:chess_component} is adjacent to a pair of cells $ \ell \in R, \ell' \in R' $ where $ \sigma(\ell^*) \ne \sigma(\ell) $ and $ \sigma(\ell^*) \ne \sigma(\ell') $, so by \Cref{claim:chess_components_closure}, $ R=R' $ and we reach a contradiction.
	If the distance between $ R $ and $ R' $ is 1, then there is a pair of cells $ \ell \in R $ and $ \ell' \in R' $ that are adjacent.
	If $ \sigma(\ell) \ne \sigma(\ell') $, by \Cref{claim:chess_component_closure_two_cells}, $ R=R' $ and we reach a contradiction.
	If $ \sigma(\ell) = \sigma(\ell') = 1 $, then each of $ \ell $ and $ \ell' $ belongs to a state-1 \nameref*{def:mono_component}, which means that $ \ell, \ell' \in X \cap Y $ and we reach a contradiction, since, by \Cref{claim:X_and_Y_disjoint_in_temporally_periodic_configurations}, $ X \cap Y = \emptyset $.
	If $ \sigma(\ell) = \sigma(\ell') = 0 $, then, since each of $ \ell $ and $ \ell' $ belongs to a \nameref*{def:chess_component}, each is adjacent to at least two cells whose state in $ \sigma $ is 1, which means that $ \sigma'(\ell) = \sigma'(\ell') = 1 $, which implies, since $ \ell $ and $ \ell' $ are adjacent, that $ \sigma''(\ell) = \sigma''(\ell') = 1 $, and we reach a contradiction to $ \sigma $ being stable.
	Hence, the distance between $ R $ and $ R' $ is at least 2 as required.
\end{proof}
\section{Proof of the Majority structural result}\label{sec:majority_structure_proof}
In this section we prove \Cref{claim:majority_strongly_stable_configurations}, the structural characterization of the stable configurations of the Majority rule on the two-dimensional torus.
In \Cref{sec:majority_structure_auxiliary}, we prove auxiliary claims and in \Cref{sec:majority_structure_proof_directions} we provide the proof of each of the statement's two directions.

\subsection{Auxiliary claims}\label{sec:majority_structure_auxiliary}

\begin{claim}\label{claim:no_monochromatic_toggling_adjacent_cells}
	Let $ \sigma: \torus{m}{n} \to \bitset $ be a stable configuration under the majority rule.
	If a pair of adjacent cells $ \ell_1, \ell_2 \in \torus{m}{n} $ are toggling in $ \sigma $ and satisfy $ \sigma(\ell_1) = \sigma(\ell_2) $, then $ \ell_1 $ and $ \ell_2 $ belong to a width-$ 2 $ \nameref*{def:zebra} $ Z \subseteq \torus{m}{n} $ in $ \sigma $, every cell $ \ell' \in \Gamma_{\le 2}(Z) \setminus Z $ is toggling in $ \sigma $ and satisfies $ \sigma(\ell') \ne \sigma(\ell'') $ where $ \ell'' \in Z \cap \Gamma_{\le 2}(\ell') $.
\end{claim}
\begin{proof}
	Without loss of generality, assume that the cells $ \ell_1 $ and $ \ell_2 $ belong to the same row $ r \in \cycnums{m} $ such that $ \ell_1 = (r,c) $ and $ \ell_2 = (r,c+1) $ for some $ c \in \cycnums{n} $.
	Let $ \sigma' = \maj(\sigma) $.
	
	We prove by induction that for every $ i \ge 0 $, the cells $ (r+i, c-1), (r+i, c), (r+i, c+1) $ and $ (r+i, c+2) $ are toggling in $ \sigma $, satisfying $ \sigma((r+i,c)) = \sigma((r+i,c+1)) $, $ \sigma((r+i,c-1)) = \sigma((r+i,c+2)) $ and $ \sigma((r+i,c)) \ne \sigma((r+i,c-1)) $, and also that if $ i $ is even, $ \sigma((r+i,c)) = \sigma((r,c)) $, and that if $ i $ is odd, $ \sigma((r+i,c)) \ne \sigma((r,c)) $, thereby establishing the claim.
	
	For $ i=0 $, the induction's base case, by the premise of the claim we have that $ \sigma((r,c)) = \sigma((r,c+1)) $ and that both $ (r,c) $ and $ (r,c+1) $ are toggling in $ \sigma $.
	Thus, it must also be the case that for every cell $ \ell_0 \in \Gamma_{=1}(\set{(r,c), (r,c+1)}) $, the cell $ \ell_0 $ satisfies $ \sigma(\ell_0) \ne \sigma((r,c)) $.
	Hence, since $ \sigma'((r,c)) = \sigma'((r,c+1)) = 1 - \sigma((r,c)) $ and the cells $ (r,c) $ and $ (r,c+1) $ are toggling in $ \sigma' $ as well, it must also be the case that for every cell $ \ell_0 \in \Gamma_{=1}(\set{(r,c), (r,c+1)}) $, the cell $ \ell_0 $ satisfies $ \sigma'(\ell_0) = \sigma((r,c)) $.
	The base case follows.
	
	For the induction step, we assume the claim holds for some $ i \ge 0 $ and show that it holds for $ i+1 $ as well.
	By the induction hypothesis, the cells $ (r+i, c) $ and $ (r+i, c+1) $ are toggling in $ \sigma $ and satisfy $ \sigma((r+i,c)) = \sigma((r+i,c+1)) = \beta $, where $ \beta = \sigma((r,c)) $ if $ i $ is even and $ \beta = 1-\sigma((r,c)) $ if $ i $ is odd.
	
	Hence, any cell $ \ell_0 \in \Gamma_{=1}(\set{(r+i,c), (r+i,c+1)}) $ satisfies $ \sigma(\ell_0) = 1-\beta $, and, in particular, $ \sigma((r+i+1,c)) = \sigma((r+i+1,c+1)) = 1-\beta $.
	Since $ \sigma'((r+i,c)) = \sigma'((r+i,c+1)) = 1 - \beta $ and the cells $ (r+i,c) $ and $ (r+i,c+1) $ are toggling in $ \sigma' $, it must be the case that for every cell $ \ell_0 \in \Gamma_{=1}(\set{(r+i,c), (r+i,c+1)}) $, the cell $ \ell_0 $ satisfies $ \sigma'(\ell_0) = \beta $.
	Thus, in particular, the cells $ (r+i+1,c) $ and $ (r+i+1,c+1) $ are toggling.
	Hence, every cell $ \ell_0 \in \Gamma_{=1}(\set{(r+i+1,c), (r+i+1,c+1)}) $ satisfies $ \sigma(\ell_0) = \beta $ and $ \sigma'(\ell_0) = 1-\beta $, and, in particular, the cells $ (r+i+1,c-1) $ and $ (r+i+1,c+2) $ are toggling and satisfy $ \sigma((r+i+1,c-1)) = \sigma((r+i+1,c+2)) = \beta $.
	
	The claim follows.
\end{proof}

\subsection{The Proof}\label{sec:majority_structure_proof_directions}
\Cref{claim:majority_strongly_stable_configurations} follows from \nameCref{claim:majority_strongly_stable_configurations_first_direction}s \ref{claim:majority_strongly_stable_configurations_first_direction} (first direction) and \ref{claim:majority_strongly_stable_configurations_second_direction} (the second direction), stated next.

\begin{claim}\label{claim:majority_strongly_stable_configurations_first_direction}
	If a configuration $ \sigma : \torus{m}{n} \to \bitset $ admits a partition $ \mathcal{P} $ that satisfies requirements \ref{req:components}-\ref{req:zebra} of \Cref{claim:majority_strongly_stable_configurations}, then the configuration $ \sigma $ is stable.
\end{claim}
\begin{proof}
	Let $ \sigma' = \maj(\sigma) $ and $ \sigma'' = \maj(\sigma') $.
	We need to show that $ \sigma(\ell) = \sigma''(\ell) $ for every cell $ \ell \in \torus{m}{n} $.
	
	Let  $ \mathcal{P} = \bigcup_{\beta=0}^3 \mathcal{P}_{\beta} $ be a partition of $ \torus{m}{n} $ that satisfies the requirements.
	For every $ \beta \in \set{0,1,2,3} $, let $ P_{\beta} = \bigcup_{C \in \mathcal{P}_{\beta}} C $.
	We prove that every cell $ \ell \in P_0 \cup P_1 $ is fixed and that every cell $ \ell \in P_2 \cup P_3 $ is toggling, thereby establishing the claim.
	
	Let $ \ell \in \torus{m}{n} $ be any cell and let $ \beta \in \set{0, 1} $ be the cell's state in $ \sigma $, i.e., $ \sigma(\ell) = \beta $.
	We claim that if $ \ell \in P_0 \cup P_1 $, then $ \sigma'(\ell) = \beta $ and that if $ \ell \in P_2 \cup P_3 $, then $ \sigma'(\ell) = 1-\beta $.
	
	Suppose first that $ \ell \in P_3 $ and let $ C \in \mathcal{P}_3 $ be the \nameref*{def:zebra} to which $ \ell $ belongs.
	By requirement~\ref{req:zebra}, the cell $ \ell $ is adjacent to exactly one cell $ \ell' \in \torus{m}{n} \setminus C $ such that $ \sigma(\ell') \ne \sigma(\ell) $, and, by the definition of \nameref*{def:zebra}, the cell $ \ell $ is also adjacent to exactly two cells $ \ell_1, \ell_2 \in C $ such that $ \sigma(\ell_1) = \sigma(\ell_2) \ne \sigma(\ell) $.
	Thus, $ \sigma'(\ell) \ne \sigma(\ell) $.
	
	Suppose next that $ \ell \in P_2 $ and let $ C \in \mathcal{P}_2 $ be the chessboard set to which the cell $ \ell $ belongs.
	By requirement~\ref{req:chess}, the cell $ \ell $ is adjacent to at most two cells $ \ell' \notin C $ and at most one of these cells satisfies $ \sigma(\ell') = \beta $.
	Additionally, since $ C $ is a chessboard set, every cell $ \ell' \in C $ to which the cell $ \ell $ is adjacent satisfies $ \sigma(\ell') = 1-\beta $.
	Thus, as the cell $ \ell $ is adjacent to exactly 4 cells, and at most one of them is of state $ \beta $ in $ \sigma $, it must be the case that $ \sigma'(\ell) = 1-\beta $ as required.
	
	Suppose now that $ \ell \in P_1 \cup P_2 $ and let $ C \in \mathcal{P}_1 \cup \mathcal{P}_2 $ be the monochromatic set to which the cell $ \ell $ belongs.
	The number of cells outside of $ C $ to which $ \ell $ is adjacent is between 0 and 4.
	If $ \ell $ is adjacent to at most two cells outside of $ C $, then it is adjacent to at least two cells $ \ell_1, \ell_2 \in C $, and since $ C $ is monochromatic, $ \sigma(\ell_1) = \sigma(\ell_2) = \sigma(\ell) = \beta $, and therefore, in this case, $ \sigma'(\ell) = \beta $.
	If the cell $ \ell $ is adjacent to three cells $ \ell' \notin C $, then, by requirement~\ref{req:mono}, at least one of them satisfies $ \sigma(\ell') = \beta $, and if $ \ell $ is adjacent to four cells $ \ell' \notin C $, then, again by requirement~\ref{req:mono}, at least two of them satisfy $ \sigma(\ell') = \beta $.
	In any case, the cell $ \ell $ is adjacent to at least two cells satisfying $ \sigma(\ell') = \sigma(\ell) = \beta $, and hence $ \sigma'(\ell) = \beta $, as required.
	
	We have shown that every cell $ \ell \in P_0 \cup P_1 $ satisfies $ \sigma'(\ell) = \sigma(\ell) $ and that every cell $ \ell \in P_2 \cup P_3 $ satisfies $ \sigma'(\ell) \ne \sigma(\ell) $.
	We next show that every cell $ \ell \in P_0 \cup P_1 $ satisfies $ \sigma''(\ell) = \sigma'(\ell) $ and that every cell $ \ell \in P_2 \cup P_3 $ satisfies $ \sigma''(\ell) \ne \sigma'(\ell) $.
	To show this, we claim that the partition $ \mathcal{P} $ satisfies requirements \ref{req:components}-\ref{req:zebra} not only with respect to the configuration $ \sigma $ but also with respect to $ \sigma' $, and then, by applying the same argument for $ \sigma' $ instead of for $ \sigma $, we establish the claim.
	
	\begin{enumerate}
		\item \emph{Requirement~\ref{req:components}:} First, since every set $ C \in \mathcal{P}_{\beta} $ for $ \beta \in \bitset $ is a $ \beta $-monochromatic set in $ \sigma $, and $ \sigma'(\ell) = \sigma(\ell) $ for every $ \ell \in P_0 \cup P_1 $, every set $ C \in \mathcal{P}_{\beta} $ is also $ \beta $-monochromatic in $ \sigma' $.
		Second, since every set $ C \in \mathcal{P}_2 $ is a chessboard set in $ \sigma $, and $ \sigma'(\ell) \ne \sigma(\ell) $ for every $ \ell \in P_2 $, every set $ C \in \mathcal{P}_2 $ is also a chessboard set in $ \sigma' $.
		Similarly, since every set $ C \in \mathcal{P}_3 $ is a \nameref*{def:zebra} in $ \sigma $, and $ \sigma'(\ell) \ne \sigma(\ell) $ for every $ \ell \in P_3 $, every set $ C \in \mathcal{P}_3 $ is also a \nameref*{def:zebra} in $ \sigma' $.
		
		\item Requirement~\ref{req:adjacency} holds by assumption regardless of the configuration under consideration, as it does not refer to the configuration.
		
		\item \emph{Requirement~\ref{req:mono}:}
		If $ C \in \mathcal{P}_0 \cup \mathcal{P}_1 $, then for every cell $ \ell \in C $, if $ |\Gamma(\ell) \setminus C| = 3 $, since requirement~\ref{req:mono} holds for $ \sigma $, there are at least two cells $ \ell_0, \ell_1 \in \Gamma(\ell) \cap \bigcup_{C' \in \mathcal{P}_2} C' $ satisfying $ \sigma(\ell_0) = 0, \sigma(\ell_1) = 1 $, and since $ \sigma'(\ell_1) \ne \sigma(\ell_1) $ and $ \sigma'(\ell_2) \ne \sigma(\ell_2) $, the requirement holds in this case for $ \sigma' $ too.
		In the case where $ |\Gamma(\ell) \setminus C| = 4 $, since requirement~\ref{req:mono} holds for $ \sigma $, there are exactly two cells $ \ell_0', \ell_0'' \in \Gamma(\ell) \cap \bigcup_{C' \in \mathcal{P}_2} C' $ and exactly two cells $ \ell_1', \ell_1'' \in \Gamma(\ell) \cap \bigcup_{C' \in \mathcal{P}_2} C' $ satisfying $ \sigma(\ell_0') = \sigma(\ell_0'') = 0 $ and $ \sigma(\ell_1') = \sigma(\ell_1'') = 1 $.
		Again, since for each cell $ \ell' \in P_2 $, $ \sigma'(\ell') \ne \sigma(\ell) $, the requirement also holds for $ \sigma' $.
		
		\item Requirement~\ref{req:chess} also holds, like requirement~\ref{req:adjacency}, regardless of the configuration under consideration.
		
		\item \emph{Requirement~\ref{req:zebra}:}
		The requirement that every cell $ \ell \in C $ for $ C \in \mathcal{P}_3 $ is adjacent to a cell $ \ell' \in C' $ for some $ C' \in \mathcal{P}_2 \cup \mathcal{P}_3 \setminus \set{C} $ holds, like requirements \ref{req:adjacency} and \ref{req:chess}, regardless of the configuration under consideration.
		Since $ \ell, \ell' \in P_2 \cup P_3 $, as we have shown, it must be the case that $ \sigma'(\ell) \ne \sigma'(\ell) $ and that $ \sigma'(\ell') \ne \sigma'(\ell') $.
		Furthermore, since $ \sigma(\ell) \ne \sigma(\ell') $, it follows that $ \sigma'(\ell) \ne \sigma'(\ell') $, as required.
	\end{enumerate}
	
	Thus, every cell $ \ell \in P_0 \cup P_1 $ satisfies $ \sigma''(\ell) = \sigma'(\ell) $ and every cell $ \ell \in P_2 \cup P_3 $ satisfies $ \sigma''(\ell) \ne \sigma'(\ell) $, and hence every cell $ \ell \in P_0 \cup P_1 $ is fixed and every cell $ \ell \in P_2 \cup P_3 $ is toggling.
\end{proof}

\begin{claim}\label{claim:majority_strongly_stable_configurations_second_direction}
	If a configuration $ \sigma : \torus{m}{n} \to \bitset $ is stable, then it admits a partition $ \mathcal{P} $ of $ \torus{m}{n} $ that satisfies requirements \ref{req:components}-\ref{req:zebra} of \Cref{claim:majority_strongly_stable_configurations}.
\end{claim}
\begin{proof}
	Since $ \sigma $ is stable, every cell $ \ell \in \torus{m}{n} $ is either fixed or toggling in $ \sigma $.
	\begin{itemize}
		\item Let $ \mathcal{P}_0 \subset 2^{\torus{m}{n}} $ be the set of all maximal connected $ 0 $-monochromatic fixed sets in $ \sigma $.
		
		\item Let $ \mathcal{P}_1 \subset 2^{\torus{m}{n}} $ be the set of all maximal connected $ 1 $-monochromatic fixed sets in $ \sigma $.
		
		\item Let $ \mathcal{P}_3 \subset 2^{\torus{m}{n}} $ be the set of all the toggling \nameref*{def:zebra}s in $ \sigma $.
		
		\item Let $ \mathcal{P}_2 \subset 2^{\torus{m}{n}} $ be the set of all the maximal connected toggling sets $ C \subseteq \torus{m}{n} \setminus \bigcup_{C' \in \mathcal{P}_3} C' $.
	\end{itemize}
	We prove that the partition $ \mathcal{P} = \bigcup_{j=0}^{3} \mathcal{P}_j $ satisfies the requirements.
	
	\begin{enumerate}
		\item The sets in $ \mathcal{P}_0 $ and $ \mathcal{P}_1 $ are, respectively, $ 0 $- and $ 1 $-monochromatic by definition, and the sets in $ \mathcal{P}_3 $ are, also by definition, width-$ 2 $ \nameref*{def:zebra}s.
		To see why the sets in $ \mathcal{P}_2 $ are chessboard sets, assume towards a contradiction that there is a set $ C \in \mathcal{P}_2 $ that is not a chessboard set.
		That is, there is a pair of cells $ \ell_1, \ell_2 \in C $ such that $ \sigma(\ell_1) = \sigma(\ell_2) $.
		Since the cells $ \ell_1 $ and $ \ell_2 $ are toggling, by \Cref{claim:no_monochromatic_toggling_adjacent_cells}, they must belong to a toggling \nameref*{def:zebra}.
		However, this implies that the set $ C $ must belong to $ \mathcal{P}_3 $, and we reach a contradiction.
		Thus, every set in $ \mathcal{P}_2 $ is a chessboard set, and hence requirement~\ref{req:components} holds.
		
		\item For every $ \beta \in \set{0,1,2} $, no two sets in $ \mathcal{P}_{\beta} $ can be adjacent to each other without violating the maximality in the definition of $ \mathcal{P}_{\beta} $, and hence requirement~\ref{req:adjacency} holds.
		
		\item Let $ P_{\beta} = \bigcup_{C' \in \mathcal{P}_{\beta}} C' $ for each $ \beta \in \set{0, 1, 2} $.
		
		Suppose $ C \in \mathcal{P}_{\beta} $ for $ \beta \in \bitset $ and let $ \ell \in C $ be a cell that is adjacent to more than than two cells outside of $ C $.
		Let $ k $ be the number of cells inside $ C $ in the von-Neumann Neighborhood of $ \ell $.
		That is, $ \Gamma_{\le 1}(\ell) \cap C = k $ and $ k \in \set{1, 2} $.
		
		Since $ \ell \in P_{\beta} $ for $ \beta \in \bitset $, the cell $ \ell $ is fixed in $ \sigma $, and therefore $ \sigma'(\ell) = \beta $, so the cell $ \ell $ is adjacent to at least $ 3-k $ cells $ \ell' \notin C $ satisfying $ \sigma(\ell') = \beta $.
		These cells $ \ell' $ cannot belong to $ P_{\beta} $ because, since they are outside of $ C $, belonging to $ P_{\beta} $ is in contradiction to the maximality of $ C $.
		These cells also cannot belong to $ P_3 $, because every cell in $ P_3 $ is monochromatic and toggling, and by \Cref{claim:no_monochromatic_toggling_adjacent_cells}, such cells cannot be adjacent to a fixed cell.
		Hence, these cells must belong to $ P_2 $.
		
		Since it must also be the case that $ \sigma''(\ell) = \beta $ (again, because the cell $ \ell $ is fixed in $ \sigma $), there are at least $ 3-k $ cells $ \ell'' \notin C $ satisfying $ \sigma'(\ell'') = \beta $, and, again, these cells cannot belong to $ P_{\beta} $ (as it will be in contradiction to the maximality of $ C $), so these cells belong to $ P_2 $, and hence, for each such cell $ \ell'' $, $ \sigma(\ell'') = 1-\beta $.
		
		That is, if $ |\Gamma_{=1}(\ell) \setminus C| = 3 $, there are at least two cells $ \ell_0, \ell_1 \in \Gamma_{=1}(\ell) \cap P_2 $ satisfying $ \sigma(\ell_0) = 0, \sigma(\ell_1) = 1 $, and if $ |\Gamma_{=1}(\ell) \setminus C| = 4 $, there are exactly two cells $ \ell_0', \ell_0'' \in \Gamma_{=1}(\ell) \cap P_2 $ and exactly two cells $ \ell_1', \ell_1'' \in \Gamma_{=1}(\ell) \cap P_2 $ satisfying $ \sigma(\ell_0') = \sigma(\ell_0'') = 0 $ and $ \sigma(\ell_1') = \sigma(\ell_1'') = 1 $, and requirement~\ref{req:mono} holds.
		
		\item Let $ C \in \mathcal{P} $ be a set in $ \mathcal{P}_2 $ and suppose towards a contradiction that a cell $ \ell \in C $ is adjacent to at least two cells $ \ell_1, \ell_2 \in \bigcup_{C' \in \mathcal{P}_0} C' $.
		Thus, if $ \sigma(\ell) = 0 $, then $ \sigma'(\ell) = 0 $, but since $ C \in \mathcal{P}_2 $, the cell $ \ell $ must be toggling in $ \sigma $, so we reach a contradiction.
		If $ \sigma(\ell) = 1 $, since $ \ell $ is toggling in $ \sigma $, it must be the case that $ \sigma'(\ell) = 0 $, and since the cells $ \ell_1 $ and $ \ell_2 $ are fixed in $ \sigma $, it must be the case that $ \sigma'(\ell_1) = \sigma'(\ell_2) = 0 $, and therefore $ \sigma''(\ell) = 0 $, and we reach a contradiction to the assumption that $ \sigma $ is stable.
		Hence, the cell $ \ell $ is adjacent to at most one cell belonging to $ \bigcup_{C' \in \mathcal{P}_0} C' $, and by a similar argument, it is also adjacent to at most one cell belonging to $ \bigcup_{C' \in \mathcal{P}_1} C' $, so requirement~\ref{req:chess} holds.
		
		\item By \Cref{claim:no_monochromatic_toggling_adjacent_cells}, every cell $ \ell \in P_3 $ can only be adjacent to toggling cells.
		Hence, if a cell $ \ell' \in \torus{m}{n} $ is adjacent to $ \ell $, then $ \ell' \in P_2 \cup P_3 $.
		Additionally, if the cells $ \ell $ and $ \ell' $ do not belong to the same set $ C \in \mathcal{P}_3 $, it cannot be the case that $ \sigma(\ell) = \sigma(\ell') $, because that would imply that there are at least three cells $ \ell'' \in \Gamma_{\le 1}(\ell) $ satisfying $ \sigma(\ell'') = \sigma(\ell) $, which implies that $ \sigma'(\ell) = \sigma(\ell) $, and we reach a contradiction because $ \ell $ must be toggling.
		Hence, requirement~\ref{req:zebra} holds.
	\end{enumerate}
\end{proof}

\newpage
\bibliography{ref}

\begin{thebibliography}{BGA21}

\bibitem[BE19]{k_local}
Omri Ben-Eliezer.
\newblock Testing local properties of arrays.
\newblock In {\em 10th Innovations in Theoretical Computer Science Conference,
  {ITCS} 2019, January 10-12, 2019, San Diego, California, {USA}}, volume 124
  of {\em LIPIcs}, pages 11:1--11:20. Schloss Dagstuhl - Leibniz-Zentrum
  f{\"{u}}r Informatik, 2019.

\bibitem[BGA21]{heat2021}
Sergey Bobkov, Edward Galiaskarov, and Irina Astrakhantseva.
\newblock The use of cellular automata systems for simulation of transfer
  processes in a non-uniform area.
\newblock In {\em CEUR Workshop Proceedings}, volume 2843, 2021.

\bibitem[BKR17]{testing_pattern_freeness}
Omri Ben{-}Eliezer, Simon Korman, and Daniel Reichman.
\newblock Deleting and testing forbidden patterns in multi-dimensional arrays.
\newblock In {\em 44th International Colloquium on Automata, Languages, and
  Programming, {ICALP} 2017, July 10-14, 2017, Warsaw, Poland}, volume~80 of
  {\em LIPIcs}, pages 9:1--9:14. Schloss Dagstuhl - Leibniz-Zentrum f{\"{u}}r
  Informatik, 2017.

\bibitem[Bsh23]{Bsh}
Nader~H. Bshouty.
\newblock On property testing of the binary rank.
\newblock In {\em 48th International Symposium on Mathematical Foundations of
  Computer Science, {MFCS} 2023, August 28 to September 1, 2023, Bordeaux,
  France}, volume 272 of {\em LIPIcs}, pages 27:1--27:14. Schloss Dagstuhl -
  Leibniz-Zentrum f{\"{u}}r Informatik, 2023.

\bibitem[BY22]{yuichibook}
Arnab Bhattacharyya and Yuichi Yoshida.
\newblock {\em Property Testing: Problems and Techniques}.
\newblock Springer, 2022.

\bibitem[CCL22]{traffic2022}
Nathan Cohen, Bastien Chopard, and Pierre Leone.
\newblock Maximum traffic flow patterns in interacting autonomous vehicles.
\newblock In {\em Cellular Automata - 15th International Conference on Cellular
  Automata for Research and Industry, {ACRI} 2022, Geneva, Switzerland,
  September 12-15, 2022, Proceedings}, volume 13402 of {\em Lecture Notes in
  Computer Science}, pages 281--291. Springer, 2022.

\bibitem[CD88]{heat1988}
Bastien Chopard and Michel Droz.
\newblock Cellular automata model for heat conduction in a fluid.
\newblock {\em Physics Letters A}, 126(8-9):476--480, 1988.

\bibitem[GM13]{goles2013neural}
Eric Goles and Servet Martinez.
\newblock {\em Neural and automata networks: dynamical behavior and
  applications}, volume~58.
\newblock Springer Science \& Business Media, 2013.

\bibitem[GO81]{goles1981_period_2}
Eric Goles and Jorge Olivos.
\newblock Comportement p{\'e}riodique des fonctions {\`a} seuil binaires et
  applications.
\newblock {\em Discrete Applied Mathematics}, 3(2):93--105, 1981.

\bibitem[Gol17]{odedbook}
Oded Goldreich.
\newblock {\em Introduction to property testing}.
\newblock Cambridge University Press, 2017.

\bibitem[GR17]{GR-dyn}
O.~Goldreich and D.~Ron.
\newblock On learning and testing dynamic environments.
\newblock {\em Journal of the ACM}, 64(3):1--90, 2017.

\bibitem[GZ18]{Zehmakan2018majority}
Bernd G{\"a}rtner and Ahad~N Zehmakan.
\newblock Majority model on random regular graphs.
\newblock In {\em Latin American Symposium on Theoretical Informatics}, pages
  572--583. Springer, 2018.

\bibitem[GZ21]{Zehmakan2021majority_2d_cellular_automata}
Bernd G{\"a}rtner and Ahad~N Zehmakan.
\newblock Majority rule cellular automata.
\newblock {\em Theoretical Computer Science}, 889:41--59, 2021.

\bibitem[JdS11]{population2011}
Rosana Jafelice and Patricia~Nunes da~Silva.
\newblock Studies on population dynamics using cellular automata.
\newblock {\em Cellular Automata: Simplicity Behind Complexity}, pages
  105--130, 2011.

\bibitem[LR23]{population2023}
Philia~Christi Latue and Heinrich Rakuasa.
\newblock Analysis of land cover change due to urban growth in central ternate
  district, ternate city using cellular automata-markov chain.
\newblock {\em Journal of Applied Geospatial Information}, 7(1):722--728, 2023.

\bibitem[MY23]{testing_OR_arbitrary_graphs}
Augusto Modanese and Yuichi Yoshida.
\newblock Testing spreading behavior in networks with arbitrary topologies.
\newblock {\em CoRR}, abs/2309.05442, 2023.
\newblock To appear in ICALP 2024.

\bibitem[NR21]{testing_1d_rules}
Yonatan Nakar and Dana Ron.
\newblock Testing dynamic environments: Back to basics.
\newblock In {\em International Colloquium on Automata, Languages, and
  Programming, {ICALP} 2021}, volume 198 of {\em LIPIcs}, pages 98:1--98:20.
  Schloss Dagstuhl - Leibniz-Zentrum f{\"{u}}r Informatik, 2021.

\bibitem[NR22]{1d_maj_structure}
Yonatan Nakar and Dana Ron.
\newblock The structure of configurations in one-dimensional majority cellular
  automata: From cell stability to configuration periodicity.
\newblock In {\em Cellular Automata - 15th International Conference on Cellular
  Automata for Research and Industry, {ACRI} 2022, Geneva, Switzerland,
  September 12-15, 2022, Proceedings}, volume 13402 of {\em Lecture Notes in
  Computer Science}, pages 63--72. Springer, 2022.

\bibitem[PRS21]{PRS}
Michal Parnas, Dana Ron, and Adi Shraibman.
\newblock Property testing of the boolean and binary rank.
\newblock {\em Theory Comput. Syst.}, 65(8):1193--1210, 2021.

\bibitem[RG11]{traffic2011}
D.~A Rosenblueth and C.~Gershenson.
\newblock A model of city traffic based on elementary cellular automata.
\newblock {\em Complex Systems}, 19(4):305, 2011.

\bibitem[Tur22]{trees_turau2022}
Volker Turau.
\newblock Fixed points and 2-cycles of synchronous dynamic coloring processes
  on trees.
\newblock {\em Structural Information and Communication Complexity},
  13298:265--282, 2022.

\bibitem[Zeh20]{zehmakan2020opinion}
Ahad~N Zehmakan.
\newblock Opinion forming in {E}rd{\H{o}}s--{R}{\'e}nyi random graph and
  expanders.
\newblock {\em Discrete Applied Mathematics}, 277:280--290, 2020.

\end{thebibliography}

\newpage
\appendix
\section{A lower bound for the naive approach}\label{sec:naive}
In the introduction, we comment that the naive testing algorithm, the one based solely on the observation that a configuration is stable if and only if each individual cell in the configuration is stable, requires a sample size that grows with the configuration's size.

Recall that the algorithm, which is applicable both to the Threshold-$ 2 $ rule and to the majority rule, selects, uniformly and independently at random, a sample $ \mathcal{L} \subseteq \torus{m}{n} $ of cells.
Then, for each cell $ \ell \in \mathcal{L} $, the algorithm queries the set $ \sigma[\Gamma_{\le 2}(\ell)] $, and accepts if and only if every cell $ \ell \in \mathcal{L} $ is stable with respect to the Threshold-$ 2 $ (or the majority) rule.

In this appendix, we provide examples of configurations for which the required sample size indeed grows with the configuration size, both for the Threshold-$ 2 $ rule and for the majority rule.

\begin{figure}[H]
	\centerline{\mbox{\includegraphics[width=0.76\textwidth]{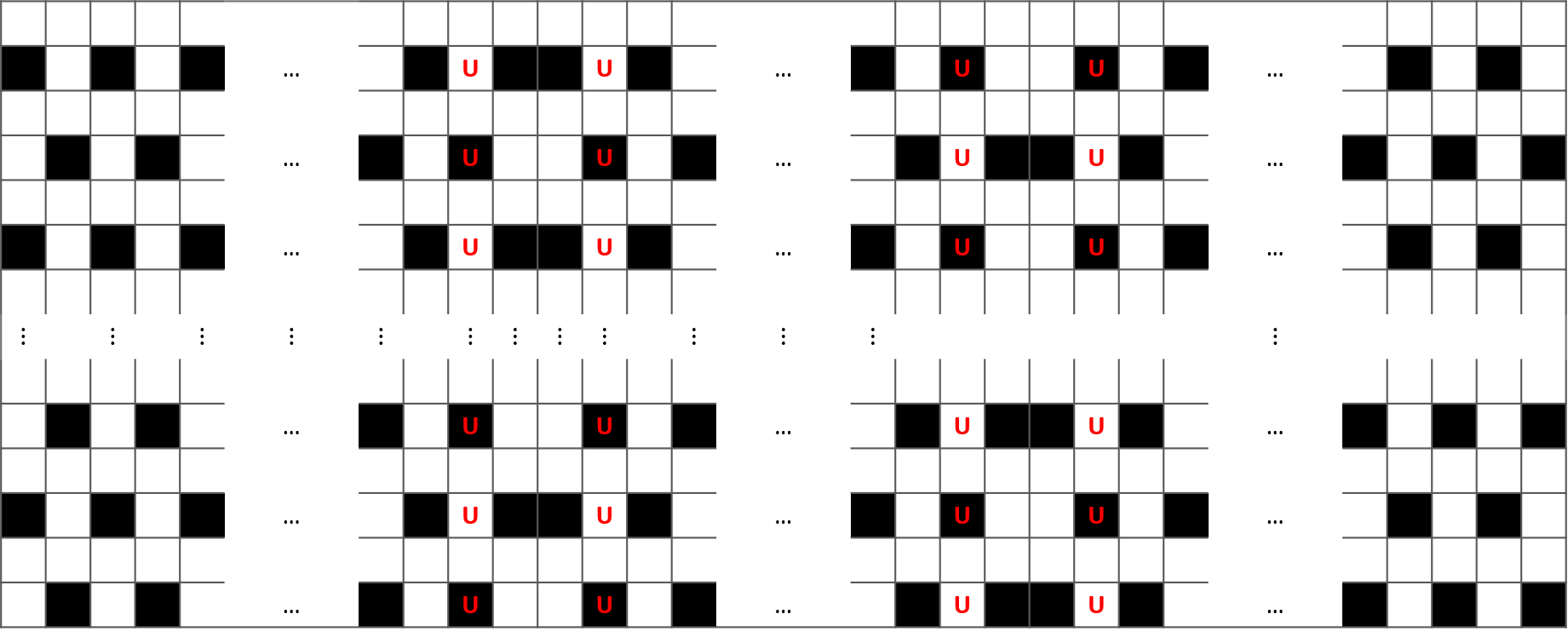}}}
	\caption{
		An example of an $ n \times n $ configuration that is $ \epsilon $-far from being stable under the Threshold-$ 2 $ rule with only $ 2n \ll \epsilon n^2 $ unstable cells, labeled by a red U.
	}
	\label{fig:thr2_hardish_instance}
\end{figure}

For the Threshold-$ 2 $ rule, assuming $ n $ is even, consider the $ n \times n $ configuration whose even rows consist entirely of zeros and whose odd rows form height-1 wraparound rectangles that are 'almost' chessboard rectangles in the following sense: they alternate between 0 and 1 until some point in the middle where a pair of adjacent cells have the same state, and then they alternate again, until a second point in the middle where a pair of adjacent cells have the same state, after which they alternate again.
See \Cref{fig:thr2_hardish_instance}.
Each such 'almost' chessboard wraparound rectangle contains a total of four unstable cells.
The configuration consists of a total of $ n / 2 $ such rows, where each pair of consecutive even rows are the `negative' of one another.
The total number of unstable cells in the configuration is only $ 2n $.
However, this configuration is $ \Omega(1) $-far from being stable under the Threshold-$ 2 $ rule.

To verify this, note that every even row consists of two incompatible chessboard parts, separated from each other on each end by a pair of cells that have the same state.
The length of each of the two parts is $ \Omega(n) $, so each part contains $ \Omega(n) $ height-$ 3 $ patterns of length $ \Theta(1) $ that are forbidden unless the row is a valid wraparound chessboard component.
Thus, to reach a stable configuration, each even row induces $ \Omega(n) $ necessary modifications, resulting in a total distance of $ \Omega(n^2) $ from every stable configuration.

\begin{figure}
	\centerline{\mbox{\includegraphics[width=0.76\textwidth]{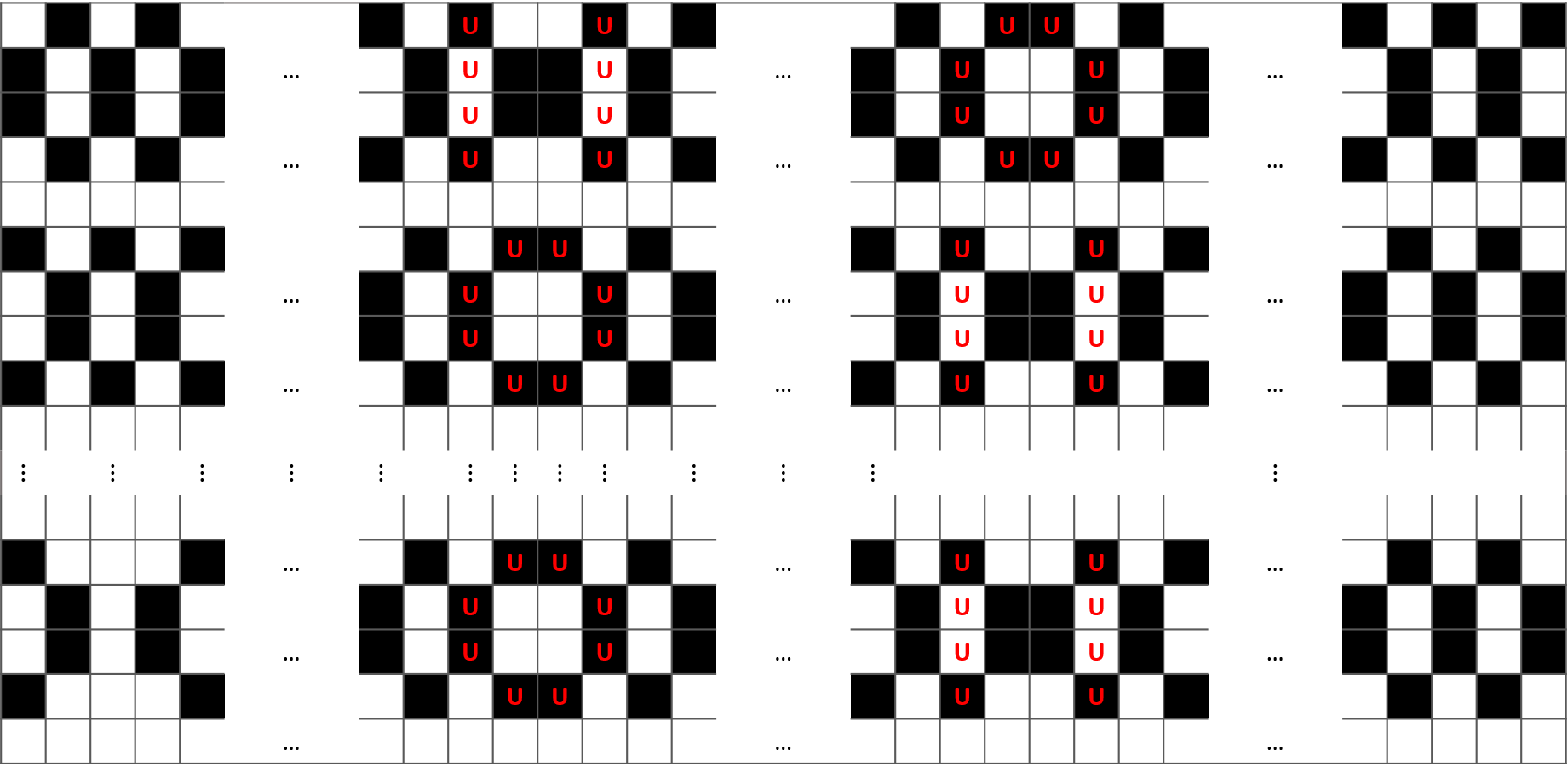}}}
	\caption{
		An example of an $ n \times n $ configuration that is $ \epsilon $-far from being stable under the majority rule with fewer than $ 4n \ll \epsilon n^2 $ unstable cells, labeled by a red U.
	}
	\label{fig:majority_hardish_instance}
\end{figure}

For the majority rule, consider the $ n \times n $ configuration that also consist of an $ n $-long structure repeating itself, but this time the structure is 4-rows high (see \Cref{fig:majority_hardish_instance}).
Analogously to the Threshold-$ 2 $ example, the configuration consists of wraparound rectangles separated by rows consisting entirely of zeros.
Each wraparound rectangle consists of a height-$ 2 $ `almost' \nameref*{def:zebra}, sandwiched between a pair of `almost' wraparound chessboard rows.
By `almost' here, we mean that there are two positions where this is violated (again, see \Cref{fig:majority_hardish_instance}), and these positions induce a total of $ 16 $ unstable cells.
Hence, the total number of unstable cells in the configuration is fewer than $ 4n $.
However, like the previous example, this configuration is $ \Omega(1) $-far from being stable under the majority rule, using an argument analogous to the one for the Threshold-$ 2 $ rule.

\end{document}